\documentclass[%
preprint,
  groupedaddress,
 amsmath,amssymb,
 aps,
 pre,
 showpacs,
]{revtex4-1}

\usepackage{graphicx}% Include figure files
\usepackage{xcolor}
\usepackage{dcolumn}% Align table columns on decimal point
\usepackage{bm}% bold math
\usepackage{psfrag}
\usepackage{dsfont}
\usepackage[export]{adjustbox}
\newcommand{\ud}{\mathrm{d}}
\providecommand{\abs}[1]{\lvert#1\rvert}
\newcommand{\bfx}{{\bf x}}
\newcommand{\bfv}{{\bf v}}
\newcommand{\bfu}{{\bf u}}

\begin{document}

%\preprint{APS/123-QED}

\title{A one-dimensional model for chemotaxis with hard-core interactions}% Force line breaks with \\

\author{Tertius Ralph}
 \email{tral001@aucklanduni.ac.nz}
\author{Stephen W. Taylor}%
\email{s.taylor@auckland.ac.nz}
\affiliation{%
 Department of Mathematics, University of Auckland, Private Bag 92019, Auckland 1142, New Zealand
}%
\author{Maria Bruna}
 \email{bruna@maths.cam.ac.uk}
 %\homepage{http://www.Second.institution.edu/~Maria.Bruna}
\affiliation{
 Department of Applied Mathematics and Theoretical Physics, Centre for Mathematical Sciences, Wilberforce Road, University of Cambridge, Cambridge CB3 0WA, United Kingdom
}%
%\affiliation{
 %Third institution, the second for Maria Bruna

\date{\today}% It is always \today, today,

\begin{abstract}
In this paper we consider a biased velocity jump process with excluded-volume interactions for chemotaxis, where we account for the size of each particle. Starting with a system of $N$ individual hard rod particles in one dimension, we derive a nonlinear kinetic model using two different approaches. The first approach is a systematic derivation for small occupied fraction of particles based the method of matched asymptotic expansions. The second approach, based on a compression method that exploits the single-file motion of hard core particles, does not have the limitation of a small occupied fraction but requires constant tumbling rates. We validate our nonlinear model with numerical simulations, comparing its solutions with the corresponding noninteracting linear model as well as stochastic simulations of the underlying particle system. 
\end{abstract}

\pacs{{05.10.Gg},{ 05.40.Fb},{ 02.30.Jr},{ 02.30.Mv}}

\maketitle

%\tableofcontents

%%%%%%%%%%%%%%%%%%%%%%%%%%%%%%%%%%%%%%%%%%%%%%
\section{Introduction} \label{sec:level1}
%%%%%%%%%%%%%%%%%%%%%%%%%%%%%%%%%%%%%%%%%%%%%%

% General bio motivation

Understanding collective dynamics and self-organization in the biological sciences has been the subject of much research interest for several decades \cite{berg1993random,couzin2002collective}. Mathematical models describing collective dynamics have been used in the context of social insects like locusts \cite{erban2011individual}, bacteria \cite{erban2004individual}, cells \cite{othmer1988models,erban2007taxis,tranquillo1987stochastic}, migratory species \cite{carrillo2014non} and robots \cite{taylor2015mathematical}. 

Models for collective dynamics can be broadly classified into three categories: particle-based models, kinetic models, and macroscopic models. Particle-based models keep track of each individual in the system explicitly, describing its motion and interactions with the others with an equation or a set of rules.
When the number of individuals is large, their behaviour is generally best studied with continuum kinetic and macroscopic models. Kinetic models consider the evolution of the density distribution of individuals in the phase space of position and velocity, whereas macroscopic models focus on the evolution of the averaged density in position only. A question of interest is how to connect the different levels of description for a given system: starting from a particle-based model, can we obtain the corresponding kinetic and/or macroscopic model? 

A large class of particle-based models in biology are the so-called velocity-jump processes, consisting of a sequence of runs and reorientations at randomly distributed times, when a new velocity is chosen \cite{othmer1988models}. Velocity-jump processes are commonly used to model the run-and-tumble dynamics of flagellated bacteria such as \emph{E. coli}, which move in a more-or-less straight line (a run) interrupted by brief tumbles  \cite{berg1993random}. Bacteria use this movement as a searching strategy: the length of the run increases and the frequency of tumbles decreases when they are moving in a favourable direction (e.g. towards food).

In its simplest form, a velocity-jump process assumes that particles move at constant speed $c$ and that tumbles or random changes in the velocity are instantaneous and  distributed according to a Poisson process of constant intensity $\lambda$. 
This velocity-jump process can be described with the kinetic equation \cite{othmer1988models}
\begin{equation}\label{CVJ}
\frac{\partial p}{\partial t} + \bfv \cdot\nabla_\bfx \,p = -\lambda\,p + \lambda\int_{V}T(\bfv,\bfu)p(\bfx,\bfu,t) \,\ud \bfu,
\end{equation}
where $p(\bfx,\bfv, t)$ is the total population density of particles located at $\bfx \in \mathbb R^d$ and moving with velocity $\bfv \in V  = \{ \bfv \in \mathbb R^d, \| \bfv \| =c \}$. Here $T(\bfv, \bfu)$ is the probability of turning from velocity $\bfu$ to velocity $\bfv$ during a tumble. Several generalisations of equation \eqref{CVJ} have been discussed in the literature, for example to include resting times \cite{othmer1988models}, account for the time particles take to turn \cite{taylor2015mathematical}, or consider  distributions other than the Poisson for the random velocity changes \cite{taylor2015birds}. See also \cite{codling2008random} for review paper on the applications of \eqref{CVJ} and related random walks in biology. It is well known that, under certain conditions on the turning kernel $T$, a parabolic scaling of space and time in  \eqref{CVJ} leads to a diffusion equation \cite{othmer2000diffusion}. In particular, if the turning kernel is unbiased (meaning that outgoing velocities are uniformly distributed on the unit sphere), the limit has isotropic diffusion with a coefficient $D = c^2/(\lambda d)$ \cite{othmer2000diffusion}. 

Equation \eqref{CVJ} can be used to describe the motion of a system of biological organisms such as cells or bacteria if interactions between them are ignored. However, a key factor that contributes to the emergence of collective behaviour in biological systems is precisely the interaction between organisms and their environments \cite{Jeckel:2019fx,Sumpter:2006br}. In the case of the velocity-jump process  \eqref{CVJ}, interactions may materialise in the turning rate $\lambda$ and/or the turning kernel $T$. For example, Erban and Othmer  \cite{erban2004individual} derive a kinetic model for bacterial chemotaxis, where $\lambda$ and $T$ depend on a chemotactic signal.  In the context of animal aggregations, Carrillo et al. \cite{carrillo2014non} consider velocity-jump processes in one and two dimensions, with interactions arising from different ``social forces'': repulsion from nearby neighbours, alignment with individuals at intermediate distances, and attraction to far-away individuals. The result is kinetic models of the form \eqref{CVJ}, but with reorientation terms $\lambda$ and $T$ being non-local functions of the density of individuals. Another example of a velocity-jump process with interactions is considered by Erban and Haskovec in \cite{erban2011individual}, to model the collective behaviour of locusts. Their model is one-dimensional and assumes that a locust switches its direction of movement with a rate that depends on the local average velocity. If most locusts nearby are moving in the opposite direction, the locust is more likely to change its direction. In both studies \cite{carrillo2014non,erban2011individual}, interactions are non-local and a mean-field limit approximation is used to obtain the kinetic models. 

The mean-field approach used in \cite{carrillo2014non,erban2011individual} is not suitable for excluded-volume or steric interactions, which are local by nature. These arise when accounting for the finite-size of organisms and prevent them from overlapping each other. 
Because of the challenges that the singular nature of the forces associated with excluded-volume interactions pose, most of the work in the literature concerns lattice-based models with simple exclusion mechanisms. These assume that individual agents occupy positions on a regular lattice and allow each lattice site to be occupied by at most a single agent. Then the lattice spacing is thought of as representing the diameter of the organism. For example, Treloar et al. \cite{treloar2011velocity} consider a one-dimensional discrete-time unbiased velocity-jump process where the move of an agent to a new position is either aborted if  it would involve stepping on other agents, or shortened to as far as it can move before colliding with another agent. By assuming that the probabilities of neighbouring sites being occupied are independent and considering an appropriate limit of the lattice spacing and time step, they obtain a system of two nonlinear kinetic equations for left- and right-moving particles. The nonlinearity appears in the flux terms and is proportional to the free or available space. Thus, in areas where agents are densely packed, the flux is reduced. A similar lattice-based velocity-jump process is considered in \cite{Slowman:2016hv} for two hard-core interacting particles. They obtain an exact expression for the stationary distribution of the two particles comprising three components: jammed (particles block each other), attractive (they move together), and independent at large separations. 

A notable exception is the work by Franz et al. \cite{franz2016hard}, which combines an unbiased off-lattice velocity-jump process with excluded-volume interactions. In particular, the authors consider a system of hard-core interacting particles evolving according to \eqref{CVJ}, but with additional changes in velocity whenever two particles collide. They assume collisions are reflective so that the particles' speeds remain constant. Starting from the $N-$particle transport equation and the BBGKY hierarchy, they derive effective transport equations for the one-particle density, still of the form \eqref{CVJ} but with a density-dependent turning rate $\lambda$. They consider two cases. For very dilute systems, they use a dilute-gas approximation in which collisions appear in the equation as a Boltzmann integral, and for more crowded systems, they use an approximation of the two-particle density adapted from the result for Brownian hard sphere particles in \cite{bruna2012excluded1}. They then obtain the diffusion limit in both cases using a moment-closure approximation, with a corresponding density-dependent collective diffusion coefficient. Interestingly, they find that the diffusion coefficient decreases as a result of collisions in the dilute-gas case, but that it increases in the crowded case for sufficiently large excluded volumes. This is in agreement with the effective diffusion coefficient derived for Brownian hard spheres in \cite{bruna2012excluded1}.

In this paper we study a one-dimensional biased velocity-jump process with excluded-volume interactions. As in \cite{franz2016hard}, we consider an off-lattice velocity-jump process and hard-core interacting particles. However, the one-dimensional problem requires a different analysis to that of the two-dimensional case studied in \cite{franz2016hard} as hard-core interactions in one dimension preserve the ordering of particles. In contrast with previous works on excluded volume \cite{franz2016hard,treloar2011velocity}, here we consider an external bias on the motion to model chemotaxis. This is achieved by allowing the turning rates to depend on the direction of motion and an external signal, and results in a drift term in the diffusive limit \cite{hillen2000hyperbolic}. 

We present two approaches to coarse-grain the particle-based model and obtain a kinetic model. The first approach is based on the method of matched asymptotic expansions, which was already used in \cite{bruna2012excluded1} in the context of Brownian particles. This approach is systematic in the limit of small volume fraction and allows us to consider spatially dependent turning rates. 
 The second approach is based on a method proposed by Rost \cite{rost1984diffusion} that exploits the fact that particles preserve the order in which they were at the initial time and is therefore heavily reliant on a one-dimensional domain. Its limitation is that the turning rates must be constant in space. This method was also used in \cite{bodnar2005derivation} in the context of Brownian particles with  interacting potentials containing hard core and a repulsive part.

The result is a set of nonlinear hyperbolic equations governing the dynamics of left- and right-moving particles. We consider the diffusion limit by taking the standard parabolic scaling. In that limit, we recover the model for hard-core Brownian particles in one dimension, known as single-file diffusion \cite{bruna2014diffusion}.
In order to validate our kinetic model, we compare it to simulations of the stochastic particle-based system under different fixed biases, as well as to the diffusion limit equation. 

The article is structured as follows. 
Section~\ref{sec:ibm} introduces the particle-level model and the equivalent $N-$particle transport equation. 
Section~\ref{sec:kinetic} is devoted to the derivation of the population-level kinetic model using the two different approaches described above. The diffusion limit and stationary solutions of the effective transport equation are presented in Section~\ref{sec:difflimit}. In
Section~\ref{sec:numerics} we present several numerical examples, comparing the solutions of our model with stochastic simulations of the particle system under different external signals and excluded volume regimes. 
We conclude in Section~\ref{sec:conclusions} with a summary and discussion of the results.

%%%%%%%%%%%%%%%%%%%%%%%%%%%%%%%%%%%%%%%%%%%%%%
\section{Individual-based model} \label{sec:ibm}
%%%%%%%%%%%%%%%%%%%%%%%%%%%%%%%%%%%%%%%%%%%%%%

We begin by describing the individual-based model in nondimensional form. We consider a group of $N$ identical hard-core particles with time-dependent positions $X_i(t)$ and velocities $V_i(t)$, $i = 1, \dots, N$. The particles are hard-rods of length $\epsilon \ll 1$ and move along a one-dimensional domain $\Omega = [0, 1]$ with no-flux boundary conditions, and move either to the left or the right with a fixed speed $c \in \mathbb R^+$, that is,
\begin{equation}
X_i(t) \in \Omega, \qquad V_i \in \{-c, c\}, \qquad \frac{\mathrm{d}X_i}{\mathrm{d}t}(t) = V_i(t).  
\end{equation}
We assume that the particles occupy a small volume fraction, so that $\epsilon N \ll 1$.

The particles undergo a velocity-jump process, where they switch their velocities to the opposite direction spontaneously based on $N$ independent Poisson processes with rates $\lambda(X_i, V_i)>0$, where
\begin{equation}
\label{switch_rates}
\lambda(x,c)=\lambda^+(x), \qquad \lambda(x,-c)=\lambda^-(x).
\end{equation}
When $\lambda^+ \ne \lambda^-$ this leads to a biased motion to one side of the domain.

Finally, our model includes hard-core interactions between particles in the following way. We assume particles switch their velocities due to collisions between each other. A particle moving right at position $X_i$ with velocity $c$ collides with a second particle at $X_i + \epsilon$ and velocity $-c$. After the collision, their velocities are reflected: the first particle has velocity $-c$ and the second particle moves at velocity $c$. Similar rules apply with the domain walls. For example, a particle moving left with velocity $-c$ will collide with the wall at $X_i = 0$ and switch its velocity to $c$ after the collision. The collisions considered are both momentum- and speed-preserving. This would not be the case in higher dimensions, where there is a distinction between elastic collisions, which preserve momentum, and reflective collisions, which preserve speed \cite{franz2016hard}. 

We note that this is a simple model for chemotaxis with steric interactions only, and that in general chemotactic cells and organisms will interact with each other and obstacles in their environment with a combination of mechanisms. For example, in the context of bacterial chemotaxis recent experimental studies have shown that cells do not always reverse their velocities upon collision with each other and boundaries (tumbling-collisions) but can also experience a so-called forward scattering \cite{Makarchuk:2019hl}. Our velocity-reversal assumption would correspond to the former case of tumbling-collisions (we point out that forward scattering is not possible for hard-core particles in one dimension). An alternative interaction rule, still accounting for the excluded volume, would have been to assume that upon collision particles block each other and remain jammed without moving, until eventually a random velocity switch frees them. This leads to an effective attraction between otherwise repulsive particles that, according to \cite{Slowman:2016hv}, could potentially explain the mechanism behind motility-induced phase separation.

%%%%%%%%%%%%%%%%%%%%%%%%%%%%%%%%%%%%%%%%%%%%%%
\subsection{Equivalent PDE description}
%%%%%%%%%%%%%%%%%%%%%%%%%%%%%%%%%%%%%%%%%%%%%%

The aim of this paper is to obtain a population-level description of the system of $N$ interacting particles. To do so, it is convenient to first write the individual-based model described above as a partial differential equation in terms of the joint probability density $P(\vec x, \vec v, t)$ in both space  $\vec x = (x_1, \dots, x_N) \in \Omega^N$ and velocity $\vec v = (v_1, \dots, v_N) \in V^N$, where $V = \{-c,c\}$, at time $t$.  It satisfies the following transport equation
\begin{equation}\label{Ndim_equation}
\frac{\partial P}{\partial t} + \vec v\cdot\nabla_{\vec x}P + \sum_{i=1}^N \left[\lambda(x_i,v_i)P(\vec x,\vec v,t)-\lambda(x_i,-v_i)P(\vec x,s_i \vec v,t) \right] = 0,
\end{equation}
where $\nabla_{\vec x} $ stands for the gradient with respect to the $N-$particle position vector $\vec x \in \Omega^N$, and $s_i$ is the operator that switches the $i$th component of $\vec v$,
\begin{equation}
\label{switch_operator}
s_i \vec v = (v_1, \dots, -v_i, \dots, v_N). 
\end{equation}
Due to the hard-core interactions between particles, \eqref{Ndim_equation} is not defined for $\vec x\in \Omega^N$, but for $\vec x \in \Omega_\epsilon^N$, where
\begin{equation}\label{Space}
\Omega_\epsilon^N=\{\vec x\in\Omega^N : |x_i-x_j|>\epsilon,\,  \forall i\ne j\}.
\end{equation}
Equation \eqref{Ndim_equation} is complemented with boundary conditions on $\partial \Omega_\epsilon^N$. The boundary condition corresponding to a collision between particles $i$ and $j$ is:
\begin{equation}\label{collision_ij}
P(\vec x, \vec v, t) = P(\vec x, s_i s_j\vec v, t), \quad \text{at} \quad |x_j - x_i| = \epsilon \quad \text{and} \quad v_i v_j<0. 
\end{equation}
The boundary condition with a wall reads:
\begin{align}\label{collision_wall}
P(\vec x, \vec v, t) &= P(\vec x, s_j \vec v, t), \quad \text{at} \quad x_j = 0, 1. 
\end{align}
We suppose that the initial positions of the particles are independent and identically distributed with initial condition
\begin{equation}
	\label{PN_0}
	P(\vec x, \vec v, 0 ) = P_0(\vec x, \vec v).
\end{equation}
This implies that $P_0$ is invariant to permutations of the particle labels and, in turn, due to the form of \eqref{Ndim_equation}, that $P$ itself is invariant to particle label permutations for all time. In particular, this means that particles are indistinguishable and their ordering (which is fixed by the initial condition due to the hard-core interactions) is not accessible/available to us.  This is important in our subsequent analysis. Finally,  $P$ satisfies the normalisation condition
\begin{equation}
\label{normalisation_P}
\int_{\Omega_\epsilon^N \times V^N} P(\vec x, \vec{v},t) \, \ud \vec x \,\ud \vec v  = 1.
\end{equation}

%%%%%%%%%%%%%%%%%%%%%%%%%%%%%%%%%%%%%%%%%%%%%%
\section{Derivation of the kinetic model} \label{sec:kinetic}
%%%%%%%%%%%%%%%%%%%%%%%%%%%%%%%%%%%%%%%%%%%%%%

Although linear, the equation \eqref{Ndim_equation} is very high-dimensional (for $N$ large) and impractical to solve directly. For this reason, we want to obtain a population-level description of the system, based on the evolution of the marginal density $p$ of one particle, say the first one,  defined as
\begin{equation}\label{marginal}
p(x_1, v_1, t) = \int_{\Omega^N_\epsilon(x_1)  \times V^{N-1} } P(\vec x,\vec v, t)  \, \ud x_2 \cdots \ud x_N \, \ud v_2 \cdots \ud v_N,
\end{equation}
where $\Omega^N_\epsilon(x_1)$ denotes the slice of configuration space $\Omega_\epsilon^N$ in \eqref{Space} for $x_1$ fixed. Since all particles are identical, the particle choice is not important. We note that, by integrating both sides of \eqref{marginal} with respect to $x_1$ and $v_1$ and using \eqref{normalisation_P}, we have that $p$ also has unit mass.

%%%%%%%%%%%%%%%%%%%%%%%%%%%%%%%%%%%%%%%%%%%%%%
\subsection{Noninteracting particles case} \label{sec:eps0}
%%%%%%%%%%%%%%%%%%%%%%%%%%%%%%%%%%%%%%%%%%%%%%

We first consider the simple case on noninteracting particles with $\epsilon =0$, so that there are no interactions between them. In one dimension one must distinguish between point particles and noninteracting particles, since point particles may still interact via hard-core interactions \cite{Ryabov:2011ix}. However, as we will discuss in Subsection \ref{sec:rost}, this distinction is only relevant for systems without invariance of collective properties under particle relabelling (for example when interested in the dynamics of an individual particle or in systems with nonidentical particles). Noninteracting particles travelling in opposite directions can pass each other and exchange order, unlike in the interacting case. Particles are therefore independent and the configuration domain has no holes. Then inserting  $P (\vec x, \vec v, t) = \prod_{i=1}^{N}p(x_{i},v_{i},t)$ in \eqref{Ndim_equation} we find that $p(x,v,t)$ satisfies
\begin{equation}\label{linear_1particle_equation}
\frac{\partial p}{\partial t}+v\frac{\partial p}{\partial x}+\lambda(x,v)p -\lambda(x,-v)p(x,-v,t)=0,
\end{equation}
with $x\in \Omega$ and $v \in V$, together with the boundary condition
\begin{equation}
p(x,v,t) = p(x,-v,t), \qquad x = 0,1,
\end{equation}
and initial condition $p(x,v,t) = p_0(x,v)$,
where $p_0$ is defined from $P_0$ using \eqref{marginal}.
We define the densities of going left and right as $\rho^-(x,t) = p(x,-c,t)$ and $\rho^+(x, t) = p(x,c,t)$, respectively. They satisfy the following system of equations 
\begin{subequations}\label{pde1}
\begin{align}
\frac{\partial \rho^+}{\partial t}+c\frac{\partial \rho^+}{\partial x}+\lambda^+(x)\rho^+ -\lambda^-(x)\rho^- & =0,  \label{pde1a}\\ 
\frac{\partial \rho^-}{\partial t}-c\frac{\partial \rho^-}{\partial x}+\lambda^-(x)\rho^- -\lambda^+(x)\rho^+ &=0,\label{pde1b}
\end{align}
\end{subequations}
with $x\in \Omega$, boundary conditions $\rho^+ = \rho^-$ at $x = 0,1$, and initial conditions $\rho^\pm(x,0) = \rho^\pm_0(x) := p_0(x,\pm c)$. Using the normalisation of $p$, we have that $\int \rho^+ \ud x$ ($\int \rho^- \ud x$) is the probability that the particle is moving right (left).  

We allow for the moment the motion of particles to be subject to a constant turning rate, that is, $\lambda^+$ and $\lambda^-$ are replaced with $\lambda_0$ in \eqref{pde1}. The probability that a particle is at $(x,t)$ is $\rho(x,t) = \rho^+(x, t) + \rho^-(x, t)$. By writing equations for $\rho$ and the flux $j(x,t) = c (\rho^+(x,t) - \rho^-(x,t))$, it can be shown that $\rho(x,t)$ satisfies the second-order PDE
\begin{equation*}
    \frac{\partial^2 \rho}{\partial t^2} + 2\lambda_0\frac{\partial \rho}{\partial t} = c^2\frac{\partial^2 \rho}{\partial x^2}, 
\end{equation*}
with initial conditions $\rho(x,0) = \rho_0^+(x) + \rho_0^-(x)$ and $\partial_t p(x,0) = c \partial_x(\rho^-_0(x) - \rho_0^+(x))$, and boundary conditions $\partial_x \rho = 0$ on $x=0,1$.
It is known as the telegrapher's equation, whose applications are discussed at length in \cite{Weiss:2002hf}, and in the context of biological transport in \cite{othmer1988models}. 

%%%%%%%%%%%%%%%%%%%%%%%%%%%%%%%%%%%%%%%%%%%%%%
\subsection{Interacting particles case via matched asymptotics} \label{sec:Asym2}
%%%%%%%%%%%%%%%%%%%%%%%%%%%%%%%%%%%%%%%%%%%%%%

When particles have a finite size $\epsilon>0$, the internal boundaries in $\Omega_\epsilon^N$ mean that particles are no longer independent. We set about the process of deriving a partial differential equation for the marginal density function $p(x_1,v_1,t)$ \eqref{marginal}. To this end, we integrate \eqref{Ndim_equation} over the remaining $N-1$ particles.

It is convenient to introduce the two-particle density
\begin{equation}\label{marginal2}
P_2(x_1,x_2,v_1,v_2,t) = \int_{\Omega^N_\epsilon(x_1,x_2) \times V^{N-2}} P(\vec x,\vec v, t)  \, \ud x_3 \cdots \ud x_N \, \ud v_3 \cdots \ud v_N,
\end{equation}
where $\Omega^N_\epsilon(x_1,x_2)$ is the configuration space available to particles $3, \dots, N$ when particles 1 and 2 are at $x_1$ and $x_2$ respectively. Then we can write $p$ as
\begin{equation}\label{marginal1}
p(x_1, v_1, t) = \int_{\Omega(x_1)}P_2(x_1,x_2,v_1,-c,t) \, \ud x_2+ \int_{\Omega(x_1)}P_2(x_1,x_2,v_1,c,t) \, \ud x_2,
\end{equation}
where $\Omega(x_1)$ the region available to a second particle at $x_2$ when the first particle is at $x_1$. Note that, since the ordering of particles is unknown, it can be that $x_2<x_1$ or $x_2>x_1$. 

Integrating \eqref{Ndim_equation} over $x_2, \dots, x_N$ and $v_2, \dots, v_N$ yields exactly
\begin{align}\label{integ1}
\begin{aligned}
0= \frac{\partial p}{\partial t} & + v_1\frac{\partial p}{\partial x_1}   + (N-1)  \sum_{v_2=\pm c} v_1 P_2 \big|_{x_2=x_1-\epsilon}^{x_2 =x_1+\epsilon} - (N-1)  \sum_{v_2=\pm c} v_2 P_2 \big|_{x_2=x_1-\epsilon}^{x_2 =x_1+\epsilon}  \\
&  + \lambda(x_1,v_1)p - \lambda(x_1,-v_1)p(x_1, -v_1,t), 
\end{aligned}
\end{align}
for $x_1\in [0,1]$. If $x_1$ is closer to the boundary than $\epsilon$, then the corresponding $P_2$ terms are set to zero. 
The first term comes from noting that the configuration domain is independent of time. The second and third terms come from integrating the transport term $v_1 \partial_{x_1} P$ and using the Leibniz rule of integration and particle relabelling. The fourth term is obtained by integrating the terms $v_i \partial_{x_i} P$ for $i=2, \dots, N$ after relabelling and using the boundary condition \eqref{collision_wall} to cancel the boundary terms at $x_2=0,1$. Finally, the last two terms are the only ones remaining from the spontaneous turning terms of \eqref{Ndim_equation} after summing $v_i$ for $i\ge 2$ over $\{-c, c\}$. Further details are given in Appendix~\ref{sec:integration}. Rearranging \eqref{integ1} we obtain
\begin{equation}\label{integ2}
\begin{split}
0= \frac{\partial p}{\partial t} &+ v_1\frac{\partial p}{\partial x_1}  + 2(N-1) v_1 \left[  P_2(x_1,x_1+\epsilon,v_1,-v_1,t) - P_2(x_1,x_1-\epsilon,v_1,-v_1,t) \right] \\
& + \lambda(x_1,v_1)p - \lambda(x_1,-v_1) p(x_1,-v_1,t).
\end{split}
\end{equation}
In equation \eqref{integ2} we see that the term involving the two-particle density $P_2$ is localized at the collision between the two particles. Below we use the method of matched asymptotic expansions to evaluate it.

%%%%%%%%%%%%%%%%%%%%%%%%%%%%%%%%%%%%%%%%%%%%%%
\subsubsection{Matched asymptotic expansions} \label{sec:MAE1}
%%%%%%%%%%%%%%%%%%%%%%%%%%%%%%%%%%%%%%%%%%%%%%

In the low-volume fraction regime under consideration ($N \epsilon \ll 1$),  three-particle (and higher) interactions are negligible compared to two-particle interactions. In particular, having fixed the first particle at $x_1$, the volume in configuration space occupied by configurations involving particle 1 being at an order $\epsilon$ distance (the range of the interaction) to another particle is $O(N\epsilon)$, while the volume of three particles within the range of interactions is $O(N^2 \epsilon^2)$. In addition, we could have the first particle close to a wall and to a second particle, and the volume of such configurations is $O(N \epsilon^2)$. Since we consider the asymptotic limit $N\epsilon \ll 1$, the leading-order contributions from the interactions are two-particle interactions, and the latter two will appear at higher order in $\epsilon$. 
Mathematically, this means that when evaluating the collision term in \eqref{integ2}, we consider $x_1$ to be far from the walls and the two-particle probability density $P_2(x_1, v_1, x_2,v_2, t)$ to be governed by the dynamics of particles 1 and 2 only, independent of the remaining $N - 2$ particles. 
For this reason, in the derivation of this section we assume that $P_2$ satisfies \eqref{Ndim_equation} with $N=2$.

As discussed above, the domain $\Omega(x_1)$ has two disjoint components, a left subinterval $[0, x_1- \epsilon)$ when $x_2<x_1$ and a right subinterval $(x_1+\epsilon, 1]$ when $x_2 > x_1$. We divide each subinterval into two regions: an \emph{inner region} when two particles are close to each other $|x_1 - x_2 | \sim \epsilon$ and an \emph{outer region} when particles are far apart, $|x_1 - x_2 | \gg \epsilon$. For ease of notation we drop the subscript $2$ in the two-particle density.

In the left and right outer regions, we define $P^l (x_1, x_2, v_1, v_2, t) = P(x_1, x_2, v_1, v_2, t)$ and $P^r (x_1, x_2, v_1, v_2, t) = P(x_1, x_2, v_1, v_2, t)$ respectively, and assume particles are independent to leading order (omitting the time variable for ease of notation),
\begin{subequations}\label{Pout}
\begin{align}\label{Pout_left}
P^l(x_1, x_2, v_1, v_2) &= q(x_1,v_1)q(x_2,v_2) + \epsilon P^l_1(x_1, x_2, v_1, v_2) + \cdots,\\
\label{Pout_right}
P^r(x_1, x_2, v_1, v_2) &= q(x_1,v_1)q(x_2,v_2) + \epsilon P^r_1(x_1, x_2, v_1, v_2) + \cdots,
\end{align}
\end{subequations}
for some functions $q$,  $P^l_1$ and $P^r_1$. Inserting the ansatz \eqref{Pout} into \eqref{Ndim_equation} with $N=2$ we find that, as in the interaction-free case, the leading-order outer $q(x_1, v_1,t)$ satisfies \eqref{linear_1particle_equation}, that is, 
\begin{equation}\label{outer_eq}
\frac{\partial q}{\partial t}+v_1\frac{\partial q}{\partial x_1}+\lambda(x_1,v_1)q -\lambda(x_1,-v_1)q(x_1,-v_1,t)=0.
\end{equation}
In the inner region we introduce the following change to inner variables $x_1 = \tilde x_1$ and $x_2 = \tilde x_1 + \epsilon \tilde x$ and we define $\tilde P(\tilde{x}_1,\tilde x,v_1,v_2,t) = P(x_1, x_2, v_1, v_2,t)$. Rewriting \eqref{Ndim_equation} with $N=2$ in terms of the inner coordinates gives
\begin{equation}
\begin{split}
0  = \epsilon\frac{\partial \tilde P}{\partial t} & + \epsilon v_1 \frac{\partial \tilde P}{\partial \tilde{x}_1} + (v_2 - v_1)\frac{\partial \tilde P}{\partial \tilde x} + \epsilon \left[ \lambda(\tilde{x}_1,v_1) + \lambda(\tilde x_1 + \epsilon \tilde x,v_2)\right] \tilde P \\ 
& - \epsilon\lambda(\tilde{x}_1,-v_1)\tilde P(\tilde{x}_1,\tilde x,-v_1,v_2) - \epsilon \lambda(\tilde x_1 + \epsilon\tilde x,-v_2)\tilde P(\tilde{x}_1,\tilde x,v_1,-v_2), \label{Pin}
\end{split}
\end{equation}
where $\tilde P$ is evaluated at $(\tilde{x}_1,\tilde x,v_1,v_2,t)$ unless explicitly written. The boundary condition \eqref{collision_ij} becomes
\begin{equation}
	\label{collision_ij_inner}
	\tilde P (\tilde x_1, \tilde x, v_1, v_2) = \tilde P(\tilde x_1, \tilde x, -v_1, -v_2), \quad \text{at} \quad \tilde x = \pm 1. 
\end{equation}
The boundary condition at the walls \eqref{collision_wall} does not appear in the inner region as it corresponds to a three-body interaction as discussed above. 
Finally, the inner solution $\tilde P$ must match with the outer solution $P^l$ as $\tilde x \to -\infty$ and $P^r$ as $\tilde x \to \infty$. Expanding \eqref{Pout} in terms of the inner variables, the matching conditions are
\begin{subequations}\label{Pout_expand}
\begin{align}
\tilde P & \sim q(\tilde x_1, v_1) q(\tilde x_1, v_2) + \epsilon \tilde x q(\tilde x_1, v_1)  \frac{\partial q}{\partial \tilde x_1}(\tilde x_1, v_2)  + \epsilon P^l_1 (\tilde x_1, \tilde x_1, v_1, v_2), &\quad &\tilde x \to -\infty, \\
\tilde P & \sim q(\tilde x_1, v_1) q(\tilde x_1, v_2) + \epsilon \tilde x q(\tilde x_1, v_1) \frac{\partial q}{\partial \tilde x_1}(\tilde x_1, v_2)  + \epsilon P^r_1 (\tilde x_1, \tilde x_1, v_1, v_2), &&\tilde x \to \infty.
\end{align}
\end{subequations}
We look for a solution of \eqref{Pin}, \eqref{collision_ij_inner}, and \eqref{Pout_expand} in the left ($\tilde x <-1$) and right  ($\tilde x >1$) subdomains in powers of $\epsilon$, $\tilde P \sim  \tilde P_0 + \epsilon\tilde P_{1} + \cdots$. The leading-order problem is, in both subdomains, 
\begin{alignat}{2}
	(v_2 - v_1)\frac{\partial \tilde P_0}{\partial \tilde x}  & = 0, & & \nonumber \\
	\tilde P_0 (\tilde x_1, \tilde x, v_1, v_2) &= \tilde P_0(\tilde x_1, \tilde x, -v_1, -v_2) &\quad &\text{at} \quad \tilde x = \pm 1, \label{inner0} \\
	\tilde P_0 &\sim q(\tilde x_1, v_1) q(\tilde x_1, v_2) & & \text{as} \quad |\tilde x| \to \infty. \nonumber
\end{alignat}
Problem \eqref{inner0} is solved, regardless of the sign of $v_1 v_2$, by
\begin{equation}\label{Pin0}
\tilde P_{0} = q(\tilde{x}_1,v_1)q(\tilde{x}_1, v_2).
\end{equation}
The $O(\epsilon)$ of \eqref{Pin} reads
\begin{align*}
0  = \frac{\partial \tilde P_{0}}{\partial t} &+ v_1\frac{\partial \tilde P_{0}}{\partial \tilde{x}_1 }  + (v_2-v_1)\frac{\partial \tilde P_{1}}{\partial \tilde x}  + \left[\lambda(\tilde{x}_1,v_1) + \lambda(\tilde{x}_1,v_2)\right] \tilde P_{0} \\&- \lambda(\tilde{x}_1,-v_1)\tilde P_{0}(\tilde{x}_1,\tilde x,-v_1,v_2)  - \lambda(\tilde{x}_1,-v_2)\tilde P_{0}(\tilde{x}_1,\tilde x,v_1,-v_2).
\end{align*}
Using that the leading-order inner solution \eqref{Pin0} satisfies \eqref{outer_eq}, the equation above simplifies to 
\begin{equation}
	\label{inner1}
	0 = (v_2 - v_1) \left[
	\frac{\partial \tilde P_{1}}{\partial \tilde x} - q(\tilde x_1,v_1)  \frac{\partial q}{\partial \tilde x_1}(\tilde x_1, v_2) \right].
\end{equation}
Let us first focus on the solution of $\tilde P_1$ for $\tilde x>1$. Integrating \eqref{inner1} and using the corresponding matching condition \eqref{Pout_expand}, we find that
\begin{equation}
	\label{Pin1}
\tilde P_1 =   \tilde x q(\tilde x_1, v_1) \frac{\partial q}{\partial \tilde x_1}(\tilde x_1, v_2)  +  P^r_1 (\tilde x_1, \tilde x_1, v_1, v_2).
\end{equation}
Equation \eqref{Pin1} already satisfies the first-order inner problem when $v_1 = v_2$. When $v_1 \ne v_2$, the boundary condition \eqref{collision_ij_inner} requires that
\begin{equation}
 q(\tilde x_1, v_1)\frac{\partial q}{\partial \tilde x_1}(\tilde x_1, -v_1) +  P^r_1 (\tilde x_1, \tilde x_1, v_1, -v_1) = q(\tilde x_1, -v_1) \frac{\partial q}{\partial \tilde x_1}(\tilde x_1, v_1)  +  P^r_1 (\tilde x_1, \tilde x_1, -v_1, v_1). \label{P1r_cond}
\end{equation}
Combining \eqref{Pin0} and \eqref{Pin1}, we find that the inner solution for $\tilde x>1$ is, to $O(\epsilon)$,
\begin{equation}
	\label{Pin_sol_right}
	\tilde P \sim  q(\tilde x_1, v_1) q(\tilde x_1, v_2)  + \epsilon q(\tilde x_1, v_1) \tilde x \frac{\partial q}{\partial \tilde x_1}(\tilde x_1, v_2) + \epsilon P^r_1 (\tilde x_1, \tilde x_1, v_1, v_2),
\end{equation}
with $P_1^r$ satisfying \eqref{P1r_cond}.
Repeating the same argument when $\tilde x<-1$, we find that the inner solution in the left subdomain is given by
\begin{equation}
	\label{Pin_sol_left}
	\tilde P \sim  q(\tilde x_1, v_1) q(\tilde x_1, v_2)  + \epsilon q(\tilde x_1, v_1) \tilde x \frac{\partial q}{\partial \tilde x_1}(\tilde x_1, v_2) + \epsilon P^l_1 (\tilde x_1, \tilde x_1, v_1, v_2),
\end{equation}
with 
\begin{align*}
    P^l_1 (\tilde x_1, \tilde x_1, -v_1, v_1) & = P^l_1 (\tilde x_1, \tilde x_1,  v_1 , -v_1) + q(\tilde x_1, -v_1) \partial_{\tilde x_1} q(\tilde x_1, v_1) - q(\tilde x_1, v_1) \partial_{\tilde x_1} q(\tilde x_1, -v_1).
\end{align*}
Finally, we can relate the left and right first-order outer solutions, $P_1^l$ and $P_1^r$ respectively, using the fact that particles are identical and indistinguishable. Suppose that we are in the right outer region (defined as $x_2>x_1$) and the particles' velocities are $v_1 = -c$ and $v_2  = c$. Then the first order outer is given by $P_1^r (x_1, x_2, -c, c)$. But this same configuration could be described as a left outer region (so that $x_2 < x_1$) with $v_1= c$ and $v_2 = -c$, so that the first-order outer is $P_1^l(x_1,x_2, c, -c)$. More generally, we have the following relation
\begin{equation}
	\label{Pl_Pr_relation}
	P_1^r (x_1, x_2, v_1, v_2) = P_1^l(x_1, x_2, v_2, v_1). 
\end{equation}

%%%%%%%%%%%%%%%%%%%%%%%%%%%%%%%%%%%%%%%%%%%%%%
\subsubsection{Evaluation of the collision terms}
%%%%%%%%%%%%%%%%%%%%%%%%%%%%%%%%%%%%%%%%%%%%%%

Now we go back to the integrated equation \eqref{integ2} and use the inner solution to evaluate the collision terms, as these correspond exactly to when the two particles are in contact and thus in the inner region. We find
\begin{equation}
\begin{split}
	P_2(x_1,x_1&+\epsilon,v_1,-v_1) - P_2(x_1,x_1-\epsilon,v_1,-v_1) = \tilde P(\tilde x_1, 1, v_1, -v_1) - \tilde P (\tilde x_1, -1, v_1, -v_1) \\
	&= \epsilon q(v_1) \partial_{\tilde x_1} q(-v_1) + \epsilon P^r_1 (\tilde x_1, \tilde x_1, v_1, -v_1) + \epsilon q(v_1) \partial_{\tilde x_1}q(-v_1) - \epsilon P^l_1 (\tilde x_1, \tilde x_1, v_1, -v_1) \\
	&= \epsilon q(v_1) \partial_{\tilde x_1}q(-v_1) + \epsilon P^r_1 (\tilde x_1, \tilde x_1, v_1, -v_1) + \epsilon q(v_1) \partial_{\tilde x_1}q(-v_1) - \epsilon P^r_1 (\tilde x_1, \tilde x_1, -v_1, v_1) \\
	&= \epsilon q(v_1) \partial_{\tilde x_1}q(-v_1) + \epsilon q(v_1) \partial_{\tilde x_1}q(-v_1) + \epsilon q(-v_1) \partial_{\tilde x_1}q(v_1)  - \epsilon q(v_1) \partial_{\tilde x_1}q(-v_1) \\
	& = \epsilon\partial_{\tilde x_1} \left[ q(-v_1) q(v_1) \right],
\label{collision_term}
\end{split}
\end{equation}
where we have omitted the $\tilde x_1$ variable in $q$ for ease of notation and written $q(\tilde x_1, v_1) \equiv q(v_1)$. In the second line we have used the inner region solutions \eqref{Pin_sol_right} and \eqref{Pin_sol_left}. In the third line we have used \eqref{Pl_Pr_relation} to write $P_1^l$ in terms of $P_1^r$, and in the fourth line we have used \eqref{P1r_cond}. 

Now we use the normalisation condition on $P_2$ to determine  the outer function $q$. We find that $q(x_1, v_1, t) = p(x_1, v_1, t) + O(\epsilon)$. Writing \eqref{collision_term} in terms of $p$ and inserting it into \eqref{integ2} we find that the density $p(x,v,t)$ satisfies, to $O(\epsilon)$, the following nonlinear kinetic equation:
\begin{equation}\label{nonlinear_1particle_eqn}
\frac{\partial p}{\partial t} + v\frac{\partial p}{\partial x}  + 2 \epsilon v (N-1)\frac{\partial}{\partial x}\left[p\,p(x,-v,t)\right]  + \lambda(x,v)p - \lambda(x,-v)p(x,-v,t) = 0,
\end{equation}
where $p = p(x, v, t)$ unless explicitly given. Equation \eqref{nonlinear_1particle_eqn} can also be written in terms of the left- and right-moving densities $\rho^\pm(x,t)$ (as previously defined in Section~\ref{sec:eps0}):
\begin{subequations}\label{both-w}
\begin{align}\label{w+closedN}
\frac{\partial \rho^{+}}{\partial t} &+ c \frac{\partial }{\partial x}\left[(1 + 2\epsilon (N-1)\rho^{-}) \rho^{+} \right] + \lambda^{+}(x)\rho^{+} - \lambda^{-}(x)\rho^{-} = 0,\\
\label{w-closedN}
\frac{\partial \rho^{-}}{\partial t} &- c \frac{\partial }{\partial x}\left[(1 + 2\epsilon (N-1)\rho^{+}) \rho^{-} \right]  + \lambda^{-}(x)\rho^{-} - \lambda^{+}(x)\rho^{+} = 0.
\end{align}
\end{subequations}
In \eqref{both-w} we have only included the leading-order nonlinear term due to steric effects. There will be correction terms of $O(\epsilon^2N)$ due to higher-order terms in the two particle inner solution $\tilde P \sim \tilde P_0 + \epsilon\tilde P_{1} + \epsilon^2\tilde P_{2} + \cdots$, as well as new inner regions where three particles $O(\epsilon^2N^2)$, or two particles and the boundary $O(\epsilon^2N)$, are close. The most important of these corrections is that due to interactions between three (or more) particles. Because our asymptotic expansion is systematic, these correction terms could in principle be calculated.

We can compare \eqref{both-w} with the system considered in \cite{erban2011individual}, also for left- and right-moving interacting particles in a one-dimensional domain. While in our system interactions are due to direct collisions between particles, in \cite{erban2011individual} interactions were introduced in the switching rates so that particles tend to switch more in the direction of movement of the ensemble of particles. Accordingly, the resulting system of PDEs, obtained via a mean-field limit, is nonlinear in the reaction terms, rather than in the transport term as in \eqref{both-w}. This is also the case of the system considered in \cite{carrillo2014non}.

Another interesting comparison is that of \eqref{both-w} with the result of \cite{treloar2011velocity}, also describing a one-dimensional velocity jump process with excluded-volume interactions, but on a lattice. They consider three different cases of interactions, all of which ensure that a particle cannot move to a new lattice site if it is already occupied (the particle either aborts the move or shortens it to avoid other particles). A key difference with our model is that interactions in \cite{treloar2011velocity} are one-sided: the particle that attempted the move changes its behaviour if that would lead to a collision with a second particle, but not vice-versa. In contrast, in our model a collision leads to both particles reversing their directions. The result of this microscopic difference in the macroscopic model is that, while both models' nonlinear flux terms depend on the densities, in \eqref{both-w} they are increasing functions of the opposite-moving density whereas in \cite{treloar2011velocity} they are decreasing functions of the total density. This is similar to what happens when comparing lattice and off-lattice based models for diffusion with excluded-volume interactions: In off-lattice models, the collective diffusion of a group is enhanced with its density whereas self-diffusion (the behaviour of a tagged particle) reduces with density  \cite{Bruna:2012wu}. In the corresponding on-lattice model (with particles undergoing a simple exclusion process), the collective diffusion is proportional to the available space, therefore decreasing with the total density \cite{Simpson:2009gi}. 

%%%%%%%%%%%%%%%%%%%%%%%%%%%%%%%%%%%%%%%%%%%%%%
\subsection{Interacting particles case via compression} \label{sec:rost}
%%%%%%%%%%%%%%%%%%%%%%%%%%%%%%%%%%%%%%%%%%%%%%

In this section we will use an idea by Rost \cite{rost1984diffusion} to reduce the problem for $N$ hard-core interacting particles into a problem for point noninteracting particles. Rost's method was used for a system of Brownian hard-core particles in one dimension by Bodnar and Velazquez \cite{bodnar2005derivation}. In one dimension, the ordering of the particles is fixed by the initial conditions due to the hard-core collisions, that is, the particles cannot change order in time ($x_i < x_j$ at $t=0$ implies that $x_i < x_j$ for all times). Without loss of generality, in this section we assume that particles are labelled according to their ordering, that is, $x_1 < x_2 < \dots < x_{N-1} < x_N$. The technique by Rost is to use a coordinate change to eliminate the excluded regions between the particles. The collision boundary conditions simply state that two colliding particles exchange their velocities when they collide. Thus, a collision in the original system corresponds to a ``label swap'' in the new system. Further, the particles are indistinguishable, so the probability density functions are invariant with respect to label swaps. The ``compressed'' system of particles thus can be modelled as a system of noninteracting particles. Therefore the dynamics of a system of identical diffusing hard-core interacting rods that cannot pass each other (single-file diffusion) can be reduced to that of a system of independent point particles \cite{rost1984diffusion}. We note that this is not true if particles are distinguishable or if we are interested in the dynamics of an individual or tagged particle \cite{Ryabov:2011ix}. 

We will see that the same idea works for hard rods undergoing a velocity-jump process. In order for this method to work, we require the number of particles $N$ to be large (in contrast with the matched asymptotic expansions derivation in the previous section, where $N$ can be a small quantity), and the tumbling rates to be constants (independent of space).

\begin{figure*}
\includegraphics[width = .61\textwidth]{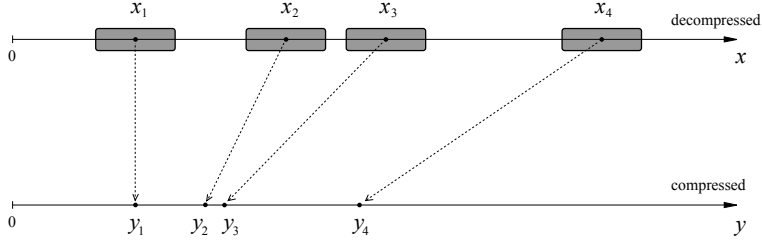}
\caption{\label{fig:scheme_compression} Sketch showing the process of compression \eqref{compression}. Modified from \cite{bodnar2005derivation}.} 
\end{figure*}

We begin by defining the change of variables to the new system. Recall that $\Omega$ is the domain available to the centres of the $N$ particles of length $\epsilon$, $x_i \in \Omega = [0, 1]$. Let us consider the position $x_i$ of the $i$th particle in the original system. We denote as $y_i$ the compressed position, which is related to $x_i$ by (see Fig.~\ref{fig:scheme_compression})
\begin{equation}
\label{compression}
y_i = x_i - (i-1) \epsilon, \quad i = 1, \dots, N. 
\end{equation} 
If $x_1 = 0$, then $y_1 = 0$, while if $x_N  = 1$ then $y_N = 1- (N-1) \epsilon$. Therefore, the domain of the compressed system is $\hat \Omega = [0, 1-(N-1) \epsilon]$. We denote time as $\hat t = t$ and define the joint probability density in the compressed system by $\hat P( \vec  y, \vec v, \hat t)$. This density takes values in $\hat \Omega^N \times V$; note that there are no ``gaps'' in this domain now.
Inserting the transformation \eqref{compression} into \eqref{Ndim_equation}, and assuming that the switching rates \eqref{switch_rates} are independent of the particle's position, we find that $\hat P$ satisfies the following equation:
\begin{equation}
\label{Ndim_equation_compressed}
\frac{\partial \hat P}{\partial \hat t}+\vec v\cdot \nabla_{\vec y} \hat P+\sum_{i=1}^N \left[\lambda(v_i)\hat P(\vec y,\vec v, \hat t)-\lambda(-v_i)\hat P(\vec y,s_i \vec v, \hat t)\right ] = 0.
\end{equation}
We now look at how the boundary conditions due to collisions between particles \eqref{collision_ij} and with the walls \eqref{collision_wall} change under the compression transformation. At the walls we have
\begin{equation}\label{collision_wall_compressed}
\hat P(\vec y, \vec v, \hat t) = \hat P(\vec y, s_i \vec v, \hat t), \quad \text{at} \quad y_i = 0, 1-(N-1)\epsilon. 
\end{equation}
For ease of exposition, we consider how the collision boundary conditions change for $N=2$. In the original coordinates, we have 
\begin{equation}\label{collision_2}
P(x_1, v_1, x_2, -v_1) = P(x_1, -v_1, x_2, v_1)
\end{equation}
at $|x_2 - x_1| = \epsilon$. Under the transformation \eqref{compression}, \eqref{collision_2} becomes
\begin{eqnarray*}\label{collision_2_compressed}
\hat P(y_1, v_1, y_2, -v_1) &=& \hat P(y_1, -v_1, y_2, v_1) \nonumber \\
&=& \hat P(y_2, -v_1, y_1, v_1) \quad \text{at} \quad  y_1 = y_2,
\end{eqnarray*}
where the last equality comes from using that $y_1 = y_2$. In general, in the $N$ particle system, a collision in the original system between two particles corresponds to swapping their labels in the compressed system. In other words, the collision boundary conditions disappear in the compressed system, and manifest instead as a label swap. In particular, this means that the compressed system will be much easier to solve since it is interaction-less, like the case we considered in Section~\ref{sec:eps0}. Finally, the initial condition is 
\begin{equation}
\label{initial_compressed}
\hat P(\vec y, \vec v,0) =  P (\vec x, \vec v,0).
\end{equation}
The problem \eqref{Ndim_equation_compressed}, \eqref{collision_wall_compressed}, \eqref{collision_2_compressed} and \eqref{initial_compressed} is separable. As a result, we can write its solution as 
\begin{equation}
\label{separable}
\hat P(\vec y, \vec v, \hat t) = \prod_{i=1}^N \hat p(y_i, v_i, \hat t),
\end{equation}
where $\hat p$ is the one-particle marginal density, defined analogously to $p$ in \eqref{marginal}. Due to the independence of particles in the compressed domain, $\hat p (y, v, \hat t)$ satisfies the same kinetic equation as in the interaction-free case \eqref{linear_1particle_equation}, that is
\begin{equation}\label{linear_1particle_equation_compressed}
\frac{\partial \hat p}{\partial \hat t}+v\frac{\partial \hat p}{\partial y}+\lambda(v)\hat p-\lambda(-v)\hat p(y,-v, \hat t)=0,
\end{equation}
with $(y, v) \in \hat \Omega \times \{ -c, c\}$, together with boundary conditions
\begin{align}
\hat p(y, -c, \hat t)  = \hat p(y, c, \hat t),\qquad x \in \partial \hat \Omega.
\end{align}
As before, we can define the two sub-densities $\hat \rho^\pm$ for the probability of going left and right, $\hat \rho^\pm (y, \hat t) = \hat p(y, \pm c, \hat t)$. These satisfy the system of equations 
\begin{subequations}
\label{compressed}
\begin{align}
\label{compressed_right}
\frac{\partial \hat \rho^+}{\partial \hat t}+c\frac{\partial \hat \rho^+}{\partial y}+\lambda^+\hat \rho^+ -\lambda^-\hat \rho^-&=0,\\
\label{compressed_left}
\frac{\partial \hat \rho^-}{\partial \hat t}-c\frac{\partial \hat \rho^-}{\partial y}+\lambda^-\hat \rho^- -\lambda^+\hat \rho^+&=0.
\end{align}
\end{subequations}
Finally, the total density of particles (moving either left or right) is given by $\hat \rho = \hat \rho^+ + \hat \rho^-$ and satisfies
\begin{equation}
	\label{eq_hat_rho}
	\frac{\partial \hat \rho}{\partial \hat t} + c\frac{\partial }{\partial y}(\hat \rho^+ - \hat \rho^-) = 0.
\end{equation}
Now we go back to the original variables and ``decompress'' the system following a procedure similar to \cite{bodnar2005derivation}. To this end, it is convenient introduce the left- and right-moving number densities $n^\pm = N \rho^\pm$, $\hat n^\pm = N \hat \rho^\pm$, and the total number densities $n = N \rho$ and $\hat n = N \hat \rho$, where $\rho = (\rho^+ + \rho^-)$. Since the macroscopic equations \eqref{linear_1particle_equation_compressed} and \eqref{compressed} are linear, the number densities $\hat n$ and $\hat n^\pm$ satisfy exactly the same corresponding equations. We note also that the number densities in the compressed system, $\hat n$ and $\hat n^\pm$ are larger than the ones in the uncompressed system, $n$ and $n^\pm$, because the same number of particles fit in a smaller region. Let's consider a small region of length $\ud x \ll 1$. Then assuming that $N$ is large such that $\ud x \ll 1/N$, the number of particles (moving either left or right) in the original system in a region of size $\ud x$ is $n (x, t)\,\ud x$ \cite{bodnar2005derivation}. The length in the compressed system where these particles are is given by 
\begin{equation} 
\ud y=(1-\epsilon n)\,\ud x. \label{distance_measure}
\end{equation}
Therefore, using that $\hat n \,\ud y = n \,\ud x$, we find
\begin{equation}\label{relationps}
\hat n = \frac{n \,\ud x}{\ud x - \epsilon n \,\ud x} = \frac{n}{1-\epsilon n }.
\end{equation}
Arguing similarly, we have
\begin{equation} \label{relationws}
\hat n^\pm = \frac{n^\pm}{1-\epsilon n}.
\end{equation}
The idea now is to use \eqref{distance_measure} to transform the equations in the compressed system to corresponding equations in the decompressed system. The original variables are related to the ones in the compressed system by
\begin{equation}
x=y+\epsilon\int_0^{y}\hat n(z,t')\,\ud z, \qquad t=\hat t. \label{variable_change}
\end{equation}
The variable change \eqref{variable_change} induces a transformation of derivatives
\begin{align}\label{chain_rule}
\frac{\partial}{\partial y} =(1+\epsilon \hat n)\frac{\partial}{\partial x}, \ \ \quad \frac{\partial}{\partial \hat t} =\frac{\partial}{\partial t}-\epsilon c(\hat n^+-\hat n^-)\frac{\partial}{\partial x},
\end{align}
where to compute $\partial_{\hat t}$ above we have used
$$\frac{\partial x}{\partial \hat t}=\epsilon\int_0^{y}\frac{\partial\hat n}{\partial \hat t}\,\ud z=-\epsilon c(\hat n^+-\hat n^-).$$
This relation is obtained from \eqref{eq_hat_rho} and the boundary condition $\hat n^+(0,\hat t)=\hat n^-(0,\hat t)$.

Applying the transformation \eqref{chain_rule} to \eqref{compressed} and inserting \eqref{relationws}, we arrive at a system of equations for the uncompressed densities $n^\pm$ of the form
\begin{align}
    {\bf A}  \begin{pmatrix} \partial_t n^+ \\ \partial_t n^- \end{pmatrix} + {\bf B}  \begin{pmatrix} \partial_x n^+ \\  \partial_ x n^- \end{pmatrix} + {\bf C}  \begin{pmatrix} n^+ \\ n^- \end{pmatrix} = 0,
\end{align}
where ${\bf A}, {\bf B}$ and $\bf C$ are $2\times2$  matrices that depend on $n^\pm$, $\epsilon$, and $\lambda^\pm$. Inverting the matrix $\bf A$ (which is invertible for $\epsilon n <1$), we obtain after some algebraic manipulation the following
\begin{subequations}\label{decomp_num0}
\begin{align}
\label{decomp_num_right0}
\frac{\partial n^+}{\partial t}+c\frac{\partial}{\partial x}\left[ \left (1 +\frac{2\epsilon n^-}{1-\epsilon n} \right) n^+  \right]+\lambda^+n^+ -\lambda^-n^-&=0,\\
\label{decomp_num_left0}
\frac{\partial n^-}{\partial t}-c\frac{\partial}{\partial x}\left[ \left (1 +\frac{2\epsilon n^+}{1-\epsilon n} \right) n^-  \right] +\lambda^-n^- -\lambda^+n^+&=0.
\end{align}
\end{subequations}

In order to compare with the system obtained in the previous subsection for small volume fraction using matched asymptotic expansions, we expand \eqref{decomp_num0} for small $\epsilon$ and retain all terms up to first order. Recalling that $n^\pm(x, t) = N \rho^\pm (x,t) = N p(x, \pm c, t)$, we find
\begin{subequations}\label{decomp_num}
\begin{align}
\label{decomp_num_right}
\frac{\partial \rho^+}{\partial t}+c\frac{\partial}{\partial x}\left(\rho^+ +2\epsilon N \rho^+ \rho^- \right) + \lambda^+\rho^+ -\lambda^- \rho^-&=0,\\
\label{decomp_num_left}
\frac{\partial \rho^-}{\partial t}-c\frac{\partial}{\partial x}\left(\rho^- +2\epsilon N \rho^+ \rho^- \right)+\lambda^- \rho^- -\lambda^+\rho^+&=0.
\end{align}
\end{subequations}
% These equations are equivalent to the following single equation for the density $p(x, v, t)$ in the decompressed system, recalling that $n^\pm(x, t) = N \rho^\pm (x,t) = N p(x, \pm c, t)$:
% \begin{equation}
% \begin{split}
% \frac{\partial {p}}{\partial t} & + v\frac{\partial }{\partial x}\left[ (1 +  2 \epsilon N p(x,-v,t) p\right] + \lambda(v){p} -\lambda(-v){p}(x,-v,t) = 0, \label{decomp_p_both}
% \end{split}
% \end{equation}
% where $p = p(x, v, t)$ unless explicitly given. 
As expected, the equation \eqref{decomp_num} derived using Rost's method, which assumes $N$ large, agrees with the equation derived using matched asymptotic expansions \eqref{both-w} in the limit of $N$ large and when the switching rate $\lambda$ is independent of position. 

In the compressed system, we have seen that, due to the particles being indistinguishable, they essentially pass through each other. In the uncompressed system, the particles must do the same, but each time a pair of particles with opposite velocities do this they jump a distance $\epsilon$, the particle diameter. This increases the probability flux. In particular, in the absence of any interaction, the probability flux of a right-moving particle is $c\rho^+(x,t)$ because such particles move with velocity $c$. However a right-moving particle will pass through approximately $2 c N \rho^-(x,t)$ left-moving particles per unit time, effectively increasing its speed by an amount $2 c \epsilon N \rho^-(x,t)$. Thus the probability flux for right-moving particles is approximately $c (1 + 2\epsilon N \rho^-)\rho^+$, in agreement with \eqref{decomp_num_right}.

%%%%%%%%%%%%%%%%%%%%%%%%%%%%%%%%%%%%%%%%%%%%%%
\section{Diffusion limit} \label{sec:difflimit}
%%%%%%%%%%%%%%%%%%%%%%%%%%%%%%%%%%%%%%%%%%%%%%

In this section we consider the long-time dynamics of the model by taking the parabolic limit of \eqref{both-w}. First, we rewrite the system in terms of the total density $\rho(x,t) = \rho^+(x, t) + \rho^-(x,t)$ and the flux $j(x,t) = c (\rho^+(x,t) - \rho^-(x,t))$:
\begin{subequations}\label{system_rho_j}
\begin{align}
	0 &= \frac{\partial \rho}{\partial t} + \frac{\partial  j}{\partial x},\\
	0 &= \frac{\partial j}{\partial t} + c^2 \frac{\partial \rho}{\partial x} + \epsilon(N-1) \frac{\partial }{\partial x} (c^2 \rho^2 - j^2) + c (\lambda^+ - \lambda^-) \rho + (\lambda ^+ + \lambda^-) j.
\end{align}
\end{subequations}
We suppose that the turning rates $\lambda^\pm$ are of the form 
\begin{equation}\label{lambda}
\lambda^{\pm} = \lambda_{0} \mp \frac{\partial S}{\partial x},
\end{equation}
where $\lambda_0(x) \ge 0$ is the base-line turning frequency and $S$ represents an external field that affects the behaviour of particles. A typical application is found in bacterial chemotaxis, where $S$ could represent an extracellular chemical concentration. Then \eqref{lambda} biases the random walk, so that a particle is less likely to change direction when moving in a favourable direction, that is, in the direction of increasing $S$. We assume that $|\partial_x S| < \lambda_0 \sim 1$ so that $\lambda^\pm \ge 0$ for all $x$. Given the form of \eqref{lambda}, we see that Rost's method in Subsection \ref{sec:rost} requires that $\partial_x S$ is constant, that is, it is valid only for linear gradients in the external field. 

We now consider the parabolic scaling by rescaling space and time as $x = x^*/\delta$ and $t = t^*/\delta^2$ with $\delta \ll 1$ \cite{carrillo2014non,othmer2000diffusion}. Then \eqref{system_rho_j} becomes (dropping the asterisks)
\begin{subequations} \label{system_rescaled}
\begin{align}
	0 &= \delta \frac{\partial \rho}{\partial t} + \frac{\partial  j}{\partial x},\\
	0 &= \delta^2 \frac{\partial j}{\partial t} + \delta c^2 \frac{\partial \rho}{\partial x} + \delta \epsilon(N-1) \frac{\partial }{\partial x} (c^2 \rho^2 - j^2)  - 2\delta c \frac{\partial S}{\partial x} \rho + 2 \lambda_0 j. 
\end{align}
\end{subequations}
We look for an asymptotic solution of \eqref{system_rescaled} of the form $\rho = \rho_0 + \delta \rho_1 + \cdots$ and $j = j_0 + \delta j_1 + \cdots$. The leading-order of \eqref{system_rescaled} is
\begin{subequations} \label{system_rescaled_O0}
\begin{align}
	0 &=  \frac{\partial  j_0}{\partial x},\\
	0 &=  2 \lambda_0 j_0, 
\end{align}
\end{subequations}
with trivial solution $j_0 = 0$. The order $\delta$ problem is, using $j_0$,
\begin{subequations} \label{system_rescaled_O1}
\begin{align} \label{O1_rho}
	0 &= \frac{\partial \rho_0}{\partial t} + \frac{\partial  j_1}{\partial x},\\ \label{O1_j}
	0 &= c^2 \frac{\partial \rho_0}{\partial x} +  \epsilon(N-1) \frac{\partial }{\partial x} (c^2 \rho_0^2 ) -2  c \frac{\partial S}{\partial x} \rho_0 + 2 \lambda_0 j_1. 
\end{align}
\end{subequations}
Inserting \eqref{O1_j} into \eqref{O1_rho} to eliminate $j_1$ we obtain the following drift-diffusion equation for $\rho_0$:
\begin{equation}\label{diffusion_limit}
	\frac{\partial \rho_0}{\partial t} =\frac{\partial}{\partial x} \left \{ \frac{c^2}{2\lambda_0}  [1 + 2 \epsilon (N-1) \rho_0 ] \frac{\partial \rho_0}{\partial x} - \frac{c}{\lambda_0} \frac{\partial S}{\partial x} \rho_0 \right \}.
\end{equation}
Identifying the diffusion coefficient as $D = c^2 / (2\lambda_0)$ and the drift as $f = (c/\lambda_0) \partial_x S$, equation \eqref{diffusion_limit} coincides with the nonlinear diffusion equation for a set of $N$ hard rods of length $\epsilon$ undergoing a Brownian motion with diffusion $D$ under a bias $f(x)$ in the limit of low occupied fraction (see Eq.~(20) in \cite{bruna2014diffusion}). 

The same calculation can be repeated starting from the kinetic model \eqref{decomp_num0} obtained via compression (setting $\lambda^\pm \equiv \lambda_0$), resulting in the diffusive limit
\begin{equation}\label{diffusive_compressed}
    \frac{\partial n}{\partial t} = \frac{\partial}{\partial x} \left[ \frac{D}{(1-\epsilon n)^2} \frac{\partial n}{\partial x}\right],
\end{equation}
where $D=c^2 / (2\lambda_0)$ again. 
This is exactly the equation obtained in \cite{bodnar2005derivation} for Brownian hard rods using the same compression method (see their Eq.~(30)) \footnote{We note that in \cite{bodnar2005derivation} they use $\rho$ instead of $n$ for the number density, $c$ instead of $\epsilon$ for the length of the rods, and the diffusion coefficient is $D=1$.}. We note we could still have considered a linear signal $S$ in \eqref{lambda}, leading to a constant drift term in \eqref{diffusive_compressed}. 

The steady-state $\rho_\infty$ of \eqref{diffusion_limit} can be found by solving
\begin{equation}\label{statsoln}
	\frac{c^2}{2\lambda_0} \left[ 1 + 2 \epsilon (N-1) \rho_\infty \right] \frac{\partial \rho_\infty}{\partial x} = \frac{c}{\lambda_0} \frac{\partial S}{\partial x} \rho_\infty.
\end{equation}
It is the same equation that we would find by setting $\partial_t = 0$ in the kinetic model \eqref{both-w}. This is to be expected since the diffusion model \eqref{diffusion_limit} is the long-time limit of the kinetic model \eqref{both-w} and, in particular, their steady-states should coincide.

The closed form solution of \eqref{statsoln} reads
\begin{equation}\label{little-wN}
\rho_\infty(x) = \frac{1}{2\epsilon(N-1)}W\left(2\epsilon(N-1)\exp \left( -\frac{1}{c}\left \{\int_{0}^{x} [\lambda^{+}(s) - \lambda^{-}(s)]\,\mathrm{d}s - A \right \} \right) \right),
\end{equation}
where $W(z)$ is the Lambert $W$ function and $A$ is the constant for normalisation. The turning frequencies $\lambda^{\pm}(x)$ are given by \eqref{lambda}. The Lambert $W$ function has an asymptotic power series expansion about zero and the $W(z)$ of \eqref{little-wN} can be represented in powers of $z$. In the power series we let $\epsilon \to 0$ giving the stationary solution for noninteracting particles
\begin{equation}\label{Big-W}
\rho_\infty(x) = \tilde A\exp\left(-\frac{1}{c}\int_{0}^{x}[\lambda^{+}(s) - \lambda^{-}(s)] \, \mathrm{d}s \right),
\end{equation}
where $\tilde A$ is the constant for normalisation.

%%%%%%%%%%%%%%%%%%%%%%%%%%%%%%%%%%%%%%%%%%%%%%
\section{Numerical examples} \label{sec:numerics}
%%%%%%%%%%%%%%%%%%%%%%%%%%%%%%%%%%%%%%%%%%%%%%

In order to assess the validity of our model, in this section we compare it with stochastic simulations of the original particle model. To test the importance of excluded-volume interactions, we also compare with the corresponding solutions for noninteracting particles. In all computations we consider $N=100$ individuals in the interval $\Omega = [0,1]$ moving with unit speed $c=1$. 

%%%%%%%%%%%%%%%%%%%%%%%%%%%%%%%%%%%%%%%%%%%%%%
\subsection{Stationary solutions}  \label{sec:stationary}
%%%%%%%%%%%%%%%%%%%%%%%%%%%%%%%%%%%%%%%%%%%%%%

We begin by comparing the stationary solution of  model 
\eqref{statsoln} (which is also the stationary solution of the diffusion limit) with the solution of the microscopic process. We assume the turning rates are of the form \eqref{lambda} and consider two cases for the base rate $\lambda_0$ and signal function $S$:
\begin{itemize}
	\item Case 1: We take $\lambda_0 = 2.5$ and consider the signal (see Fig.~\ref{Fig:signal}(a))
\begin{equation} \label{signal1}
S_1(x) = 1 - 2\abs{x - 0.5}.
\end{equation}

\item Case 2: We take $\lambda_{0} = 16$ and the signal function (see Fig.~\ref{Fig:signal}(b))
\begin{equation} \label{signal2}
S_2(x) = 4.77\,e^{-50\,(x \,-\, 0.5)^2} - 3.58\,e^{-25\,(x \,-\, 0.5)^2}.
\end{equation}
This signal is a simple example of a simple domain where toxins occupy the negative regions of $S$ and nutrients are to be found in the positive regions of $S$. 
\end{itemize}

\def \scl {1.2}
\begin{figure}[tbh!]
\unitlength=1cm
\psfrag{S1}[][][\scl][0]{$S_1$} \psfrag{S2}[][][\scl][0]{$S_2$}
\psfrag{x}[][][\scl]{$x$} \psfrag{a1}[l][][\scl]{(a)}
\includegraphics[width = 0.4\textwidth]{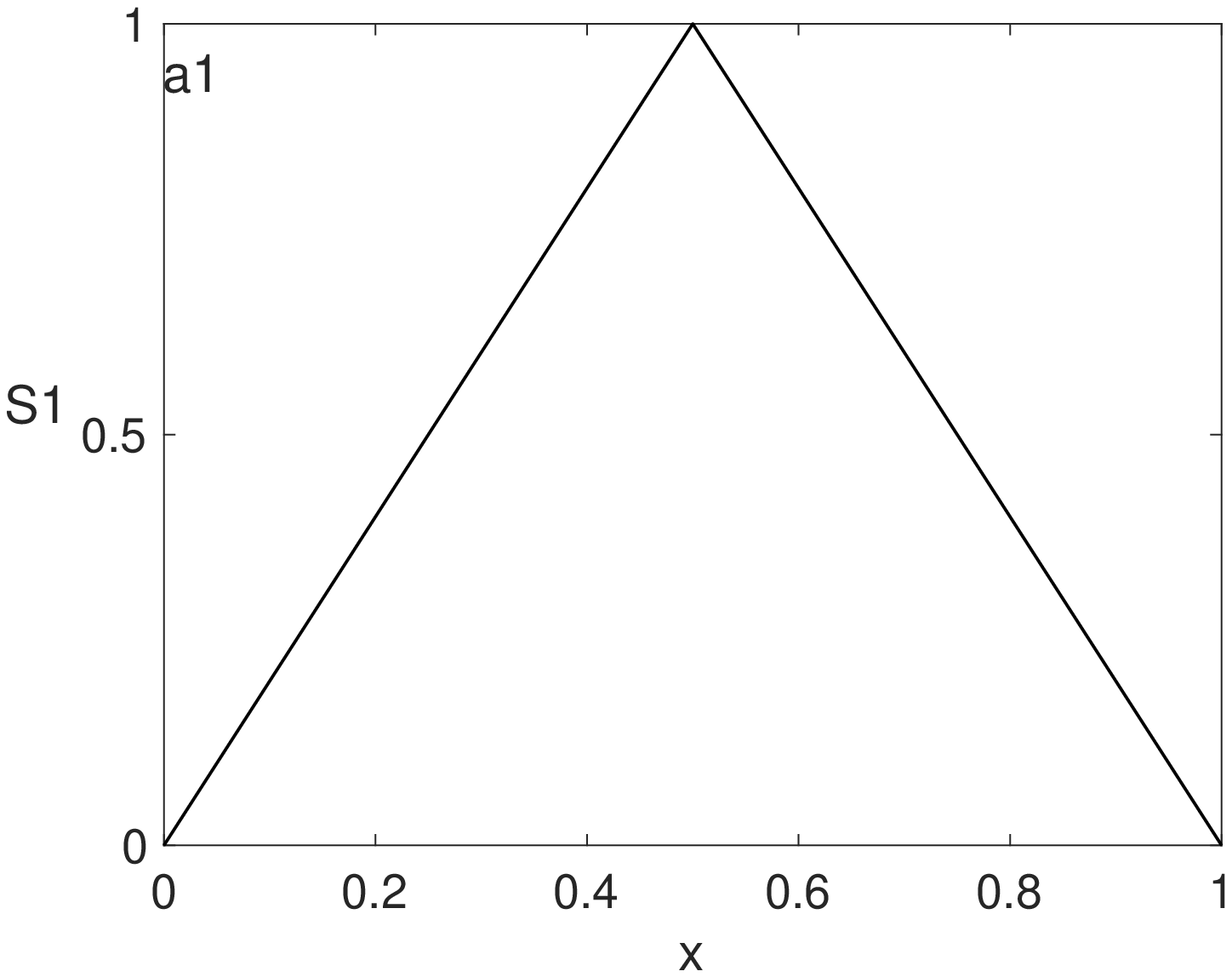} \qquad
\psfrag{w}[][][\scl]{$\rho_\infty$} \psfrag{x}[][][\scl]{$x$} \psfrag{a2}[l][][\scl]{(b)}
\includegraphics[width = 0.4\textwidth]{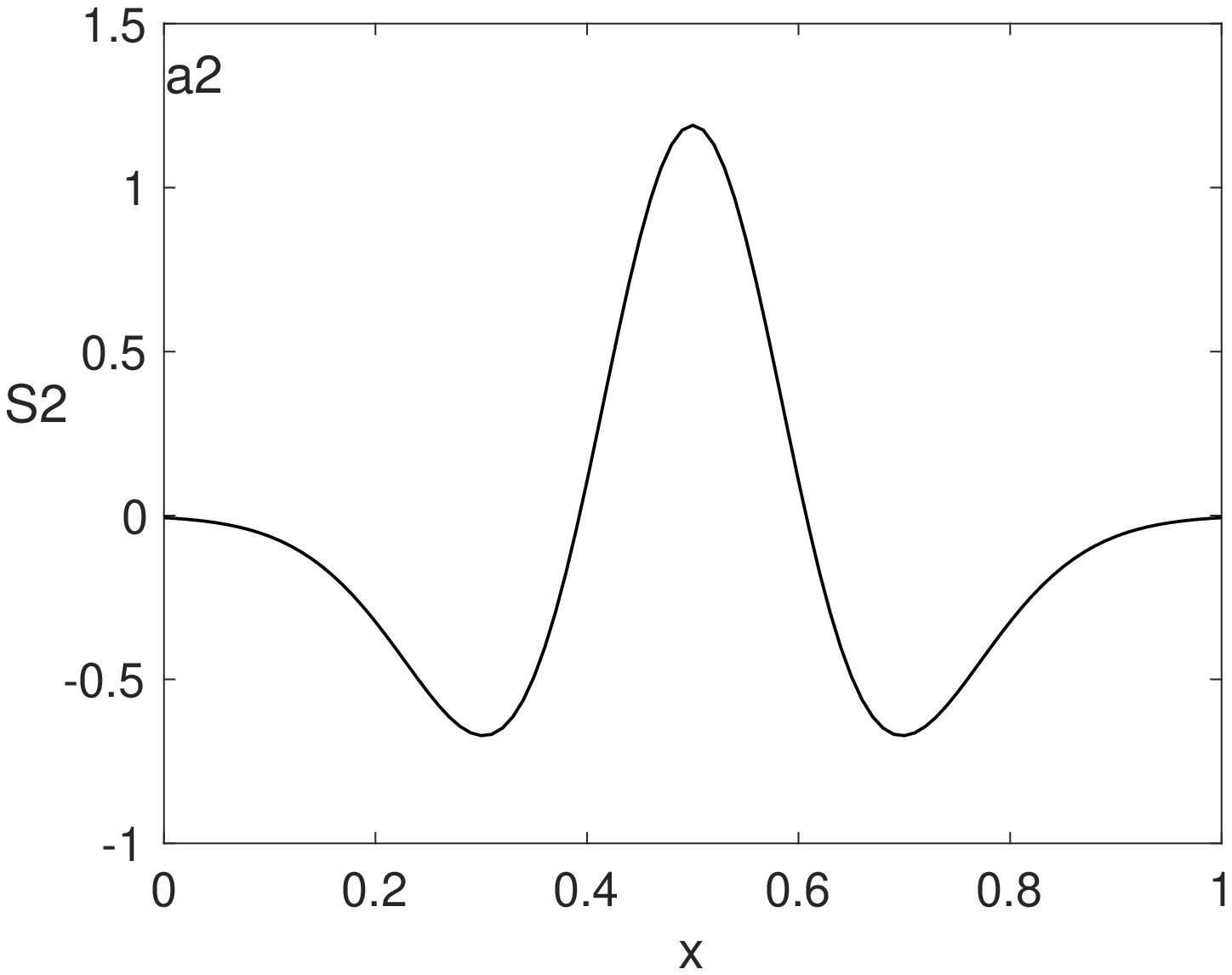} 
\caption{\label{Fig:signal} External signals $S$ used in the numerical simulations: (a) $S_1(x)$ in \eqref{signal1}, (b) $S_2(x)$ in \eqref{signal2}.}
\end{figure}

We compare the stationary densities $\rho_\infty$ predicted by \eqref{little-wN} and \eqref{Big-W} with simulations of the particle system using the Metropolis--Hastings (MH) algorithm \cite{chib1995understanding}. This algorithm allows us to sample directly from the $N-$dimensional microscopic density, which is of the Boltzmann form $C\exp(-E(\vec x))$, where $C$ is the normalisation constant and $E$ is the energy, which depends on the position of all the particles. In each step, one particle $i\in \{1,N\}$ is chosen at random and a local move from its current position $x_i$ to a new position $x'_i$ is attempted, where $x'_i = x_i + \delta \xi$,  $\xi \sim \mathcal N(0,1)$ and $\delta$ is tuned to optimise the convergence to the equilibrium distribution. The move is accepted with probability $\text{min}[1, \exp(-\Delta E)]$, where $\Delta E$ is the change in energy due to the attempted move: 
\begin{align*}
	\Delta E(\vec x, x_i') = \left\{ 
\begin{array}{l l}
(c/\lambda_0) [S(x_i) - S(x_i')], \qquad & |x_i'-x_j|\ge \epsilon, j\ne i,\\
+\infty, \qquad & \text{otherwise},
\end{array}
\right.
\end{align*}
In this way, a move is always accepted if it does not increase the energy of the system, and always rejected if it leads to two particles overlapping. 
We use $10^6$ steps of the MH algorithm for noninteracting particles and $10^7$ steps for the hard rods to generate histograms of the stationary densities. The domain is divided into 40 bins to generate the histograms.

In Fig.~\ref{Fig:MH} we show the results for both noninteracting particles and hard-core particles with the turning rates of cases 1 and 2. In Fig.~\ref{Fig:MH}(a) the rods are of size $\epsilon = 0.002$. With $N=100$ particles, this corresponds to an occupied fraction of $20\%$. 
As expected, we observe that particles aggregate around the maximum of the signal function in the centre of the domain. The interaction-free solutions within Fig.~\ref{Fig:MH} become the points of reference in allowing one to see the competition between the  most favourable signal environment and the steric repulsion in finite size particles.
The particle density around the peak of the signal functions is reduced for finite-size particles in comparison to that of noninteracting particles. This is because not all particles can be in and around the point of the maximum signal, since they would overlap each other; a redistribution occurs.  
In Fig.~\ref{Fig:MH}(b) we use instead rods of size $\epsilon = 0.001$ ($10\%$ occupied fraction) with the parameters of Case 2. 
We observe the same effect as in the previous case, namely that noninteracting particles aggregate in the centre of the domain where the maximum of the signal is, and to a lesser extent in the case of hard rods. Despite having only half of the occupied fraction in Fig.~\ref{Fig:MH}(b) relative to Fig.~\ref{Fig:MH}(a), the difference between the density profiles between noninteracting particles and rods in Case 2 is greater than in Case 1 in the central part of the domain. This is because the gradients of $S_2$ are steeper near the centre than that of $S_1$, and so in the absence of interactions particles tend to aggregate in a smaller region. Finally, we note there is a good agreement between the model predictions and the stochastic simulations in Fig.~\ref{Fig:MH}.

\def \scl {1.2}
\begin{figure}[htb]
\unitlength=1cm
\psfrag{w}[][][\scl]{$\rho_\infty$}
\psfrag{x}[][][\scl]{$x$} \psfrag{b1}[l][][\scl]{(a)}
\includegraphics[width = 0.45\textwidth]{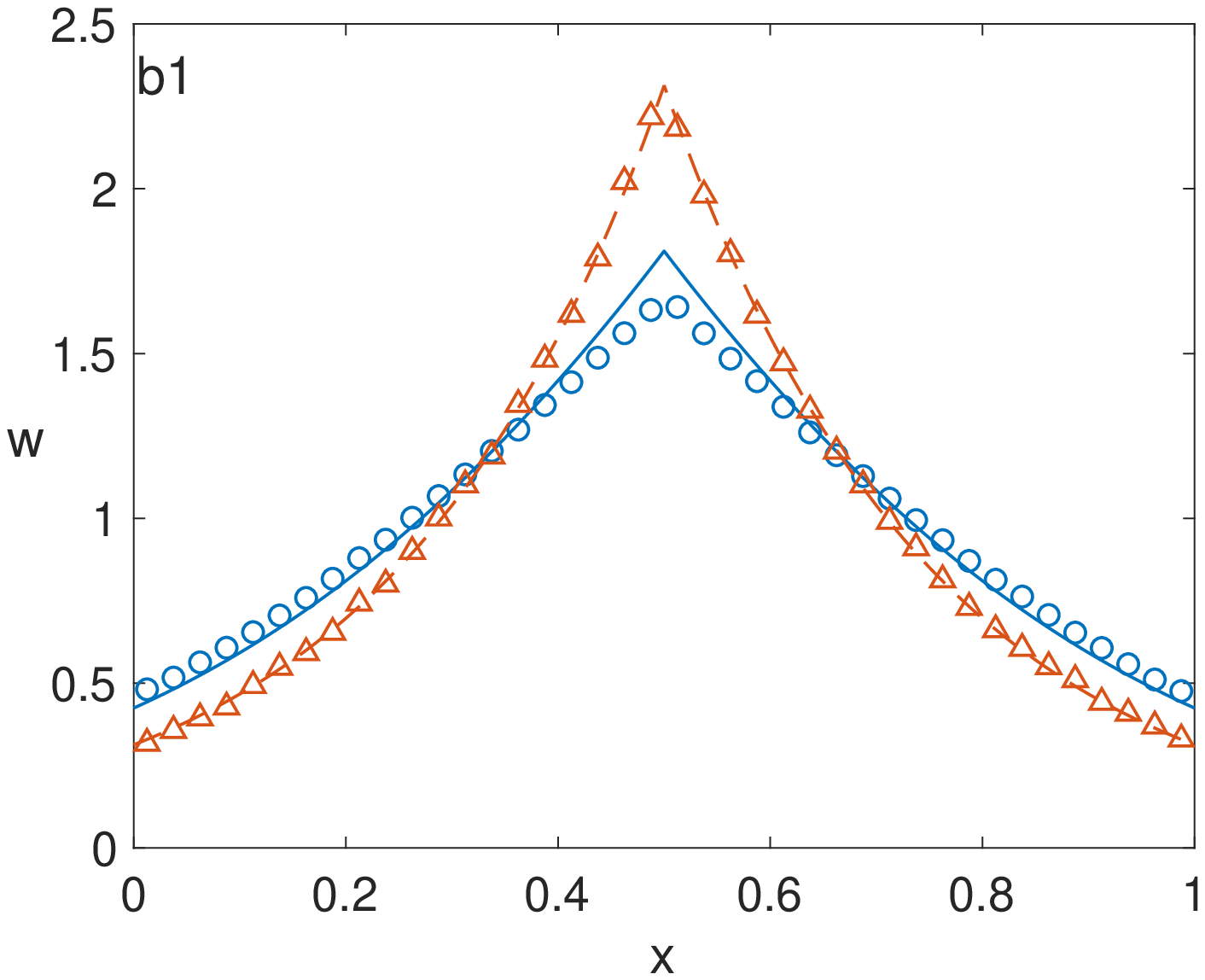}
\qquad
\psfrag{x}[][][\scl]{$x$} \psfrag{b2}[l][][\scl]{(b)}
\includegraphics[width = 0.45\textwidth]{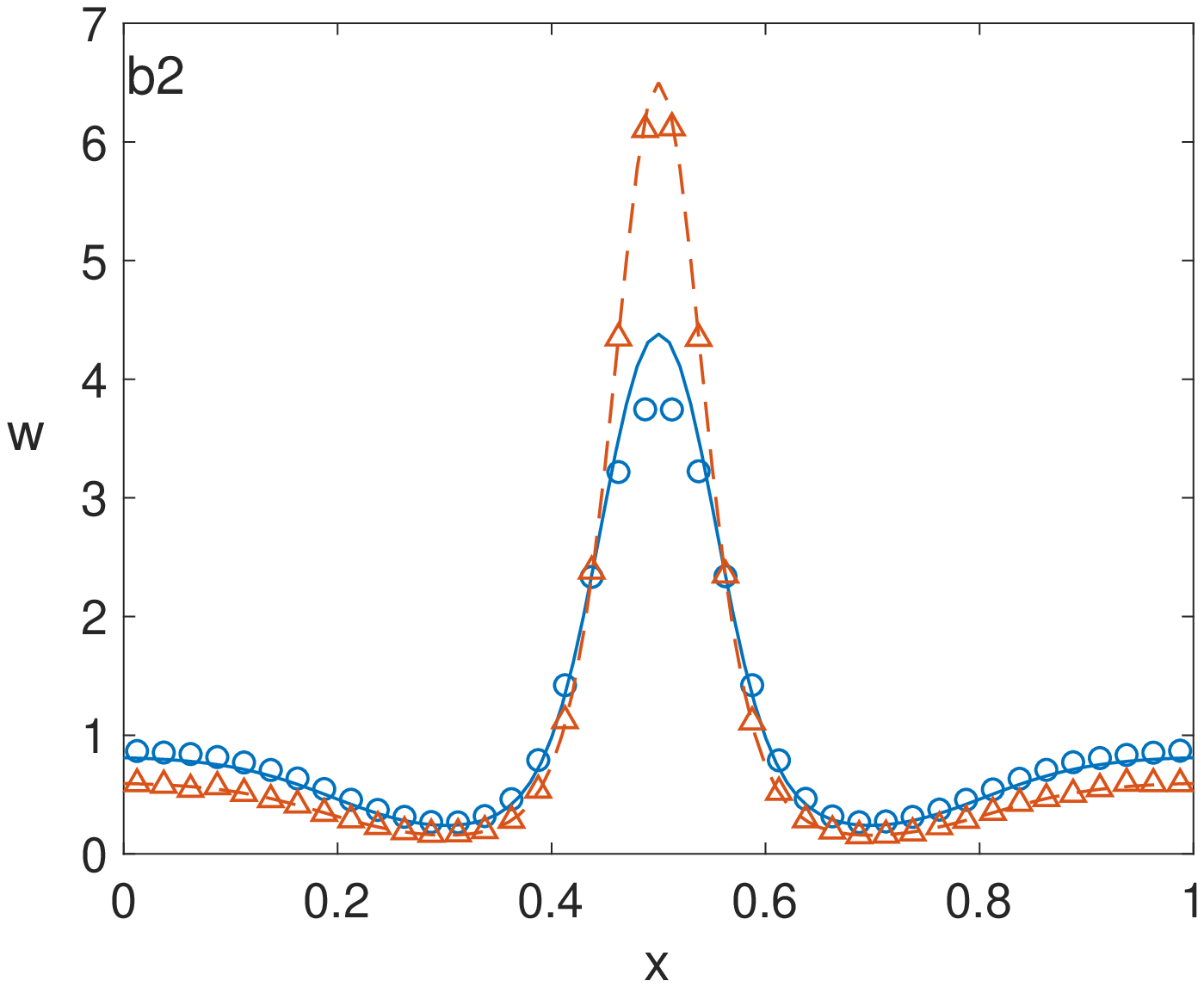}
\caption{\label{Fig:MH} 
Stationary solutions of the kinetic model for noninteracting particles \eqref{Big-W} (dashed red lines) and for hard rods \eqref{little-wN} (solid blue lines), and results of the MH simulation for noninteracting particles (red triangles) and hard rod particles (blue circles). (a) $N=100$ particles with $\epsilon = 0.002$ and signal $S_1$. MH parameter $\delta = 0.1$. (b) $N=100$ particles with $\epsilon = 0.001$ and signal $S_2$. We used $10^6$ MH steps for noninteracting particles and $10^7$ for hard rods, and steps of mean size $\delta = 0.1$ in (a) and $\delta = 0.2$ in (b).}
\end{figure}

%%%%%%%%%%%%%%%%%%%%%%%%%%%%%%%%%%%%%%%%%%%%%%
\subsection{Time-dependent simulations} \label{sec:kineticsolns}
%%%%%%%%%%%%%%%%%%%%%%%%%%%%%%%%%%%%%%%%%%%%%%

For all the computations in this subsection, we consider a set of $N=100$ particles and the initial condition:
\begin{equation} \label{initial}
	\rho^+_0(x) = (55/100)\mathds{1}_{[0.1287,0.3762]}(x), \qquad \rho_0^-(x) = (45/100)\mathds{1}_{[0.6238,0.8713]}(x).
\end{equation}

All individual-based simulations are performed using an event-based kinetic Monte Carlo (KMC) simulation of velocity-jump processes \cite{alder1959studies,lubachevsky1991simulate}. The main idea of this algorithm is that one can jump directly from one event to the other without missing events. For the simulations we divide the domain into $40$ bins. At the start of each realisation we generate random initial positions for the two groups of left- and right-moving particles in the corresponding intervals given by \eqref{initial}. To avoid the overlap of finite-size particles we exploit Rost's idea of switching to the compressed domain as seen in Fig.~\ref{fig:scheme_compression}.
We generate uniformly and independently distributed  initial conditions $Y_i(0)$ in the compressed intervals and use \eqref{compression} to establish the initial conditions  $X_i(0)$ for the hard rods  in the decompressed state.

We solve the linear kinetic system \eqref{pde1} using the method of characteristics on the transport part of the equations in tandem with integrating the source terms via the midpoint rule. We solve the nonlinear kinetic system \eqref{both-w} using the second-order Nessyahu--Tadmor (NT) central scheme \cite{nessyahu1990non} with fixed time step $\Delta t = 0.0005$ and domain divided into 404 computational cells. Part of the central scheme developed by \cite{nessyahu1990non} uses a generalized minmod limiter that includes a parameter $\theta$, which has the following range $1 \leq \theta \leq 2$. This parameter can be used to control the amount of numerical viscosity present in the resulting scheme. In all the numerical examples below, $\theta = 1.9$ is used. Comparisons against simulations on finer grids and using the Chebfun PDE solver \texttt{pde23t} \cite{driscoll2014chebfun}  showed that the simulations were sufficiently well resolved to accurately capture the wave propagation.
%We integrate the nonlinear diffusion equation \eqref{diffusion_limit} using the method of lines in tandem with MATLAB's inbuilt {\fontfamily{pcr}\selectfont ode15s} solver.

%%%%%%%%%%%%%%%%%%%%%%%%%%%%%%%%%%%%%%%%%%%%%%
\subsubsection{Transient solutions without tumbling and bias}
%%%%%%%%%%%%%%%%%%%%%%%%%%%%%%%%%%%%%%%%%%%%%%

We begin with a simple case where $\lambda^\pm \equiv 0$, that is, no random changes in the velocities of particles. This allows us to validate the numerical methods. In particular, the solution of \eqref{pde1} for noninteracting particles is simply $\rho^\pm(x,t) = \rho_0^\pm(x \mp ct)$, that is, waves travelling at constant speed right and left for $\rho^+$ and $\rho^-$ respectively. Since the only changes in velocity are due to collisions, we expect the nonlinear system \eqref{both-w} for hard rods to behave like the noninteracting particles linear system up to the point when the two fronts collide.

It is also possible to obtain an exact solution of the nonlinear kinetic model \eqref{decomp_num0} that we derived using Rost's method for the case $\lambda^\pm \equiv 0$. The system \eqref{decomp_num0} was obtained by decompressing the coordinate system from a compressed system in which the model was linear. We exploit the same transformation to solve the nonlinear system. To do this, we start by using Equation \eqref{distance_measure} and the initial number densities $n_0^\pm (x) = N \rho_0^\pm(x)$ to compress the coordinate system at time $t=0$. The initial data for the linear system is computed using \eqref{relationps}. After obtaining the exact solution of the linear system, the solution-dependent coordinate system is decompressed using \eqref{variable_change} and the solution values are obtained by inverting the relationships \eqref{relationps}.

We plot the results of the simple non-tumbling case in Figs.~\ref{fig:test_nodrift3} and \ref{fig:test_nodrift4} for $10\%$ and $20\%$ occupied fraction, respectively. We run it up to time $t=0.5$ and plot the densities $\rho^+, \rho^-$ and $\rho$ corresponding to noninteracting and hard-core particles at times $t= 0,0.1, \dots, 0.5$. As expected, the solutions of the linear system \eqref{pde1} are non-dissipative and move at a constant speed (as seen in the evolution of the means plotted in Fig~\ref{fig:means}). The solutions of the nonlinear system via matched asymptotic expansions \eqref{both-w} or Rost's methods \eqref{decomp_num0} are identical to the linear case up to $t=0.1$ (second row in  Fig.~\ref{fig:test_nodrift3}), since the two waves have just met. The collision of the waves occurs around $t=0.2$ (third row in Fig.~\ref{fig:test_nodrift3}), and we observe a deformation of the waves corresponding to hard rods.  Interestingly, after the waves have bounced off each other (from $t=0.3$) they recover the original shape and speed as precollision, albeit they are shifted outwards by a small amount. We further comment on this shift, due to the finite-size of particles, in the next section. 
\def \scl {1.2} 
\def \scp {0.9} 
\begin{figure}
\unitlength=1cm
\psfrag{x}[][][\scl]{$x$}
\psfrag{rr}[][][\scl]{Density $\rho^+$}
\psfrag{rl}[][][\scl]{Density $\rho^-$}
\psfrag{rt}[][][\scl]{Density $\rho$}
\psfrag{t0}[r][][\scp]{$t=0$}
\psfrag{t1}[r][][\scp]{$t=0.1$}
\psfrag{t2}[r][][\scp]{$t=0.2$}
\psfrag{t3}[l][][\scp]{$t=0.3$}
\psfrag{t4}[l][][\scp]{$t=0.4$}
\psfrag{t5}[l][][\scp]{$t=0.5$}
	\includegraphics[width = .9\textwidth]{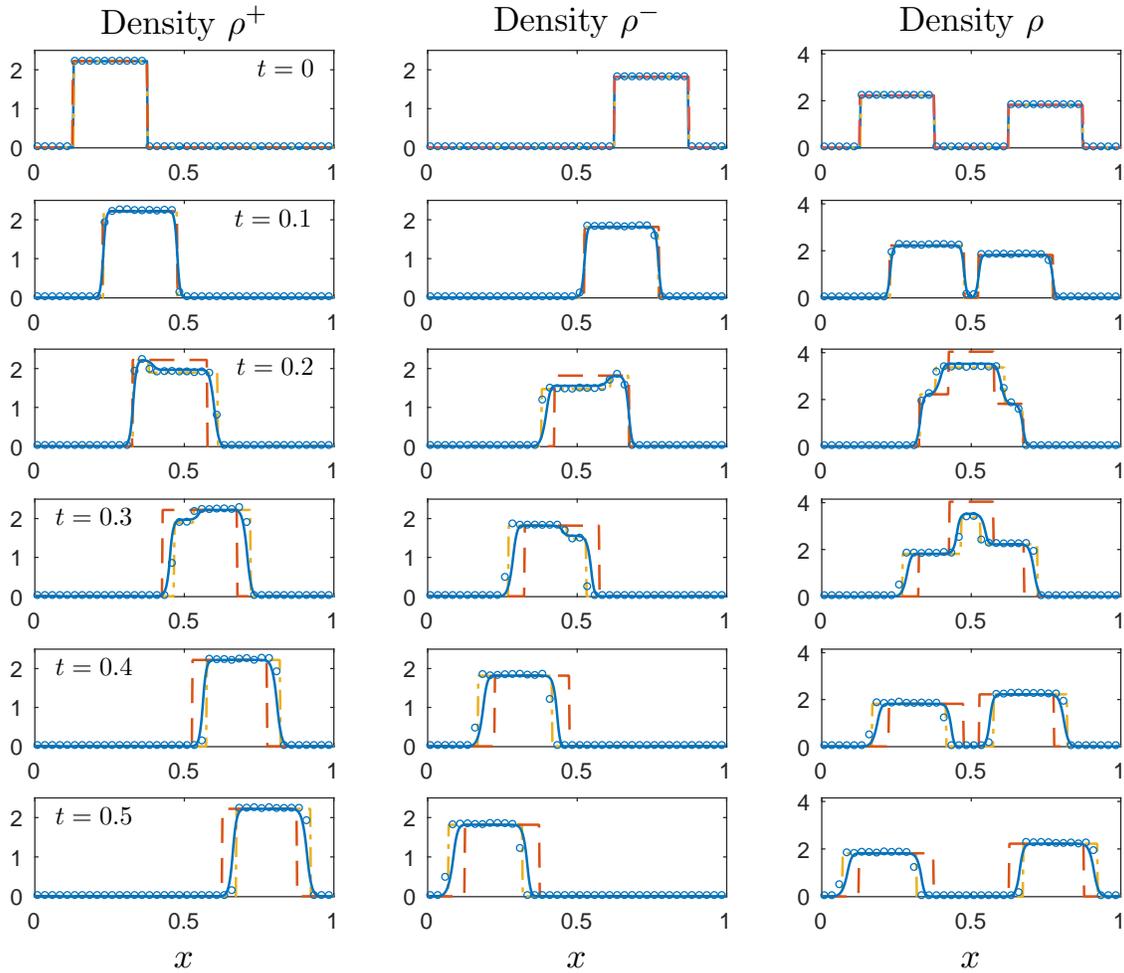}
	\caption{\label{fig:test_nodrift3} Transient solutions of the kinetic models for noninteracting particles \eqref{pde1} (dashed red lines) and hard rods via matched asymptotic expansions \eqref{both-w} (solid blue lines) and Rost's method \eqref{decomp_num0} (dot dashed yellow lines). Particle KMC simulations (blue circles) obtained  from $2.5 \times 10^3$ realisations. We used the initial condition \eqref{initial}, $\lambda^\pm = 0$, $N = 100, \epsilon = 0.001, c = 1$.}
\end{figure}
\def \scl {1.2} 
\def \scp {0.9} 
\begin{figure}
\unitlength=1cm
\psfrag{x}[][][\scl]{$x$}
\psfrag{rr}[][][\scl]{Density $\rho^+$}
\psfrag{rl}[][][\scl]{Density $\rho^-$}
\psfrag{rt}[][][\scl]{Density $\rho$}
\psfrag{t0}[r][][\scp]{$t=0$}
\psfrag{t1}[r][][\scp]{$t=0.1$}
\psfrag{t2}[r][][\scp]{$t=0.2$}
\psfrag{t3}[l][][\scp]{$t=0.3$}
\psfrag{t4}[l][][\scp]{$t=0.4$}
\psfrag{t5}[l][][\scp]{$t=0.5$}
	\includegraphics[width = .9\textwidth]{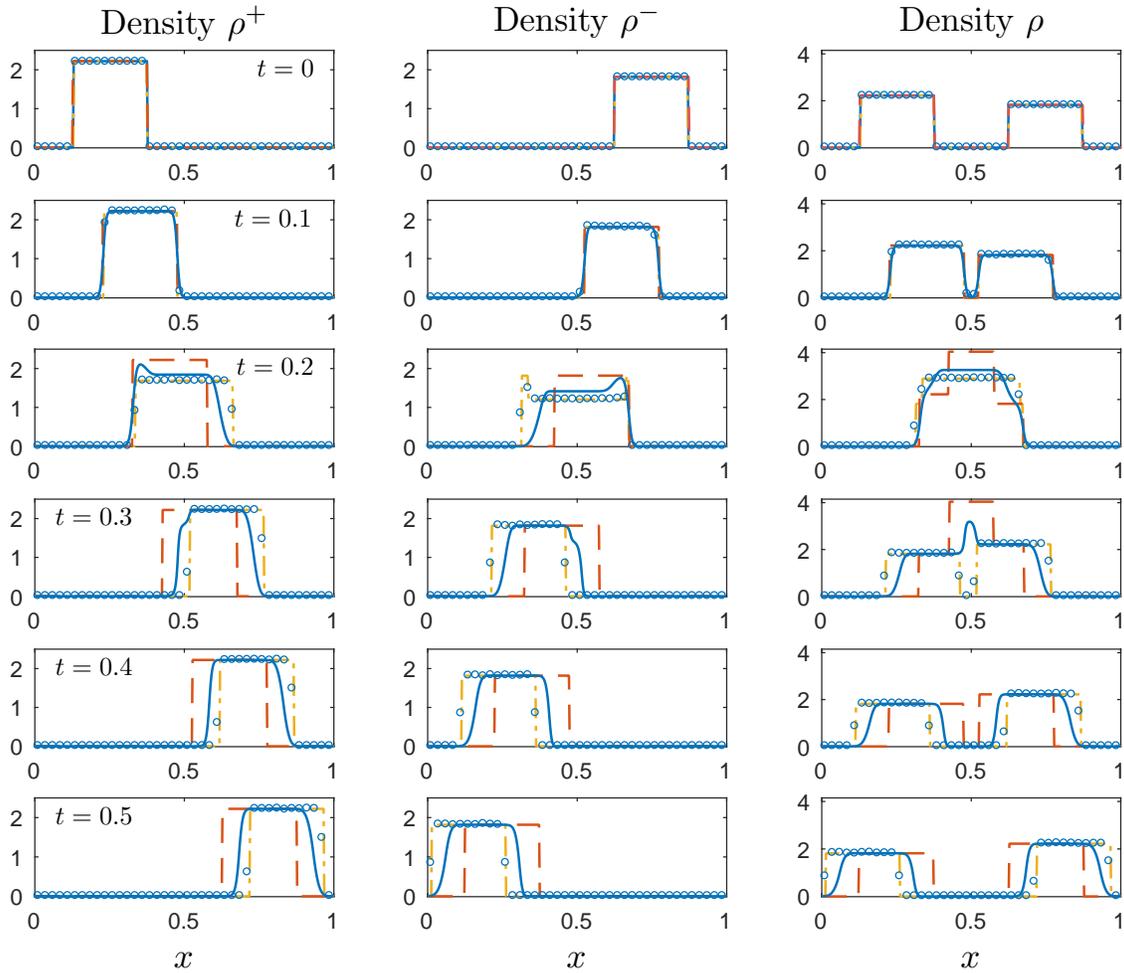}
	\caption{\label{fig:test_nodrift4} Transient solutions of the kinetic models for noninteracting particles \eqref{pde1} (dashed red lines) and hard rods via matched asymptotic expansions \eqref{both-w} (solid blue lines) and Rost's method \eqref{decomp_num0} (dot dashed yellow lines). Particle KMC simulations (blue circles) obtained  from $2.5 \times 10^3$ realisations. We used the initial condition \eqref{initial}, $\lambda^\pm = 0$, $N = 100, \epsilon = 0.002, c = 1$.}
\end{figure}

%%%%%%%%%%%%%%%%%%%%%%%%%%%%%%%%%%%%%%%%%%%%%%
\subsubsection{Wave velocities and shifts}\label{sec:Velocity}
%%%%%%%%%%%%%%%%%%%%%%%%%%%%%%%%%%%%%%%%%%%%%%

To check the speed of the waves and the effect that excluded-volume interactions have, we next compute the centre of mass $\langle \rho^\pm \rangle(t)$ of each wave at times $t = 0, 0.1, \dots, 0.5$ and plot the results in Fig.~\ref{fig:means}. For the linear system (red dashed lines) these are straight lines with slope $c = 1$ as expected. In the case of rods of length $\epsilon$, each collision causes a ``shift'' of $\epsilon$ (the distance between two particles at collision), so we expect the left- and right-moving waves to shift $(N-1)\epsilon$ with respect to the noninteracting case. This is confirmed when computing the means from the KMC simulation data (see blue circles in Fig.~\ref{fig:means}). We find that the Rost solution (yellow dot-dashed lines) is in perfect agreement with the simulations, even at the higher $20\%$ occupied fraction (Fig.~\ref{fig:means}(b)). As we increase the occupied fraction from 10 to $20\%$, the model for rods \eqref{both-w} via matched asymptotic expansions (blue solid lines) does not agree so well with simulations, since it only takes into account the first correction in volume fraction. Another contributing factor may be numerical errors with the NT scheme due to the discontinuities in the data \cite{nessyahu1990non}.
\def \scp {0.85} 
\begin{figure}
\unitlength=1cm
\psfrag{a}[][][\scl]{ (a)}
\psfrag{b}[][][\scl]{ (b)}
\psfrag{m}[][][\scl]{Centre of mass}
\psfrag{t}[][][\scl]{$t$}
\psfrag{sim}[][][\scp]{sim}
\psfrag{Rost}[][][\scp]{Rost}
\psfrag{Linear}[][][\scp]{Linear}
\psfrag{NT}[][][\scp]{NT}
	\includegraphics[height = .35\textwidth]{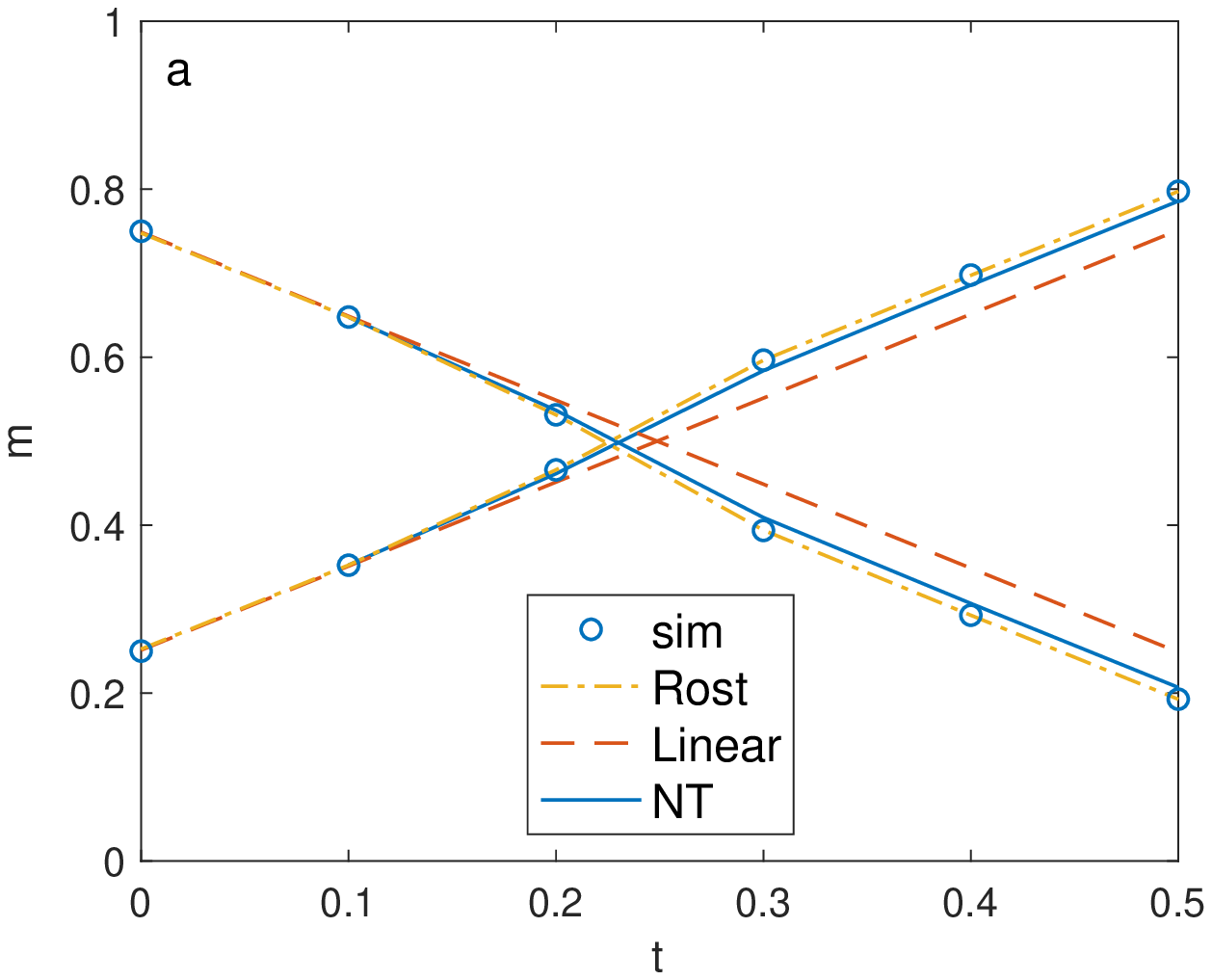} \quad \includegraphics[height = .35\textwidth]{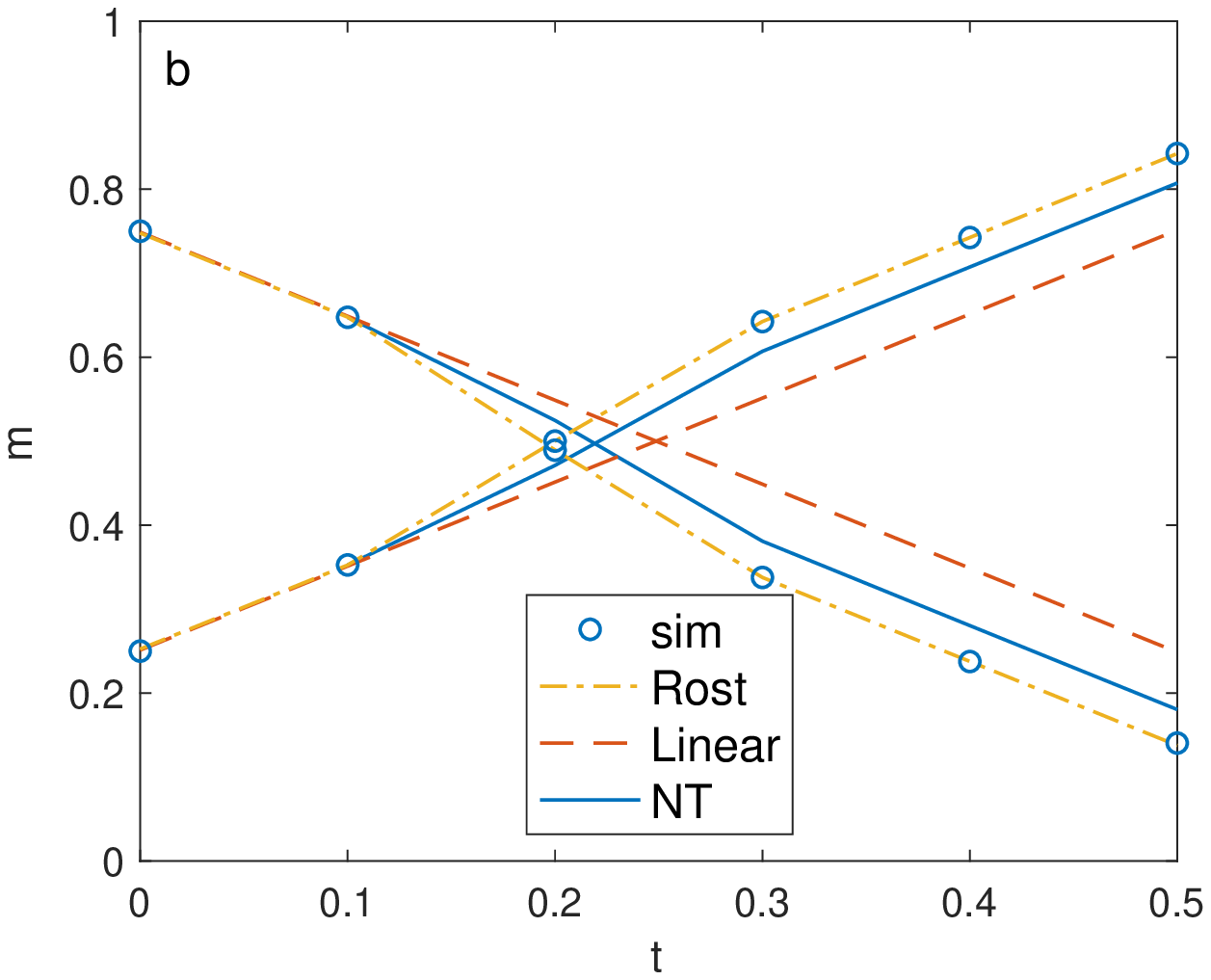}
	\caption{\label{fig:means} Centre of mass of $\rho^\pm(t)$ as a function of time for the solutions in (a) Fig.~\ref{fig:test_nodrift3} and (b)  Fig.~\ref{fig:test_nodrift4}.}
\end{figure}

We can also gain an understanding of these shifts by considering wave speeds. For the case of $\lambda^\pm \equiv 0$ under consideration, all three kinetic models, noninteracting particles \eqref{pde1}, hard rods via matched asymptotic expansions \eqref{both-w}, and hard rods via Rost's method \eqref{decomp_num0} (using $n^\pm=N\,\rho^\pm$), can be represented as a hyperbolic system
\begin{align*}\label{homogeneous}
%\begin{split}
\frac{\partial \vec{\rho}}{\partial t} + c {\bf M}\frac{\partial \vec{\rho}}{\partial x} & = \vec{0}, \qquad x \in \mathbb{R}, \qquad t > 0, \\
\vec{\rho}(x,0) & = \vec{\rho}_0(x),
%\end{split}
\end{align*}
where $\vec \rho = (\rho^+, \rho^-)^\top$ and ${\bf M}$ is a $2\times 2$ matrix. 
For the noninteracting case, ${\bf M} = \text{diag}(1,-1)$ and the equations are uncoupled with travelling wave solutions $\rho^+(x,t) = \rho_0^+(x - ct)$ and $\rho^-(x,t) = \rho_0^-(x + ct)$, each moving with speed $c$. These travelling wave solutions are plotted in Fig.~\ref{fig:test_nodrift3} as dashed red lines.

The systems \eqref{both-w} and \eqref{decomp_num0} do not decouple, but disturbances move along the characteristic curves of these equations at speeds given by the eigenvalues of $\bf M$ \cite{fritzjohn}.
For the model obtained by matched asymptotic expansions \eqref{both-w},
\begin{equation*}
{\bf M} =  \begin{pmatrix} 1 + 2\delta \rho^{-} & 2\delta \rho^{+} \\ - 2\delta \rho^{-} & -(1 + 2\delta \rho^{+}) \end{pmatrix},
\end{equation*}
where $\delta = (N-1) \epsilon$. In this case, the eigenvalues of the matrix $\bf M$ are
\begin{equation*}
\Lambda_{1}(\vec \rho) = a + \sqrt{b} \; > \; \Lambda_{2}(\vec \rho) = a - \,\sqrt{b},
\end{equation*}
where
\begin{equation*}
a =  \delta(\rho^{-} - \rho^{+}), \qquad b = 1 + 2\delta(\rho^{+} + \rho^{-}) + \delta^{2}(\rho^{+} - \rho^{-})^{2}. 
\end{equation*}
Expanding these in powers of $\delta$, we find that
\begin{subequations} \label{evalsMAE}
\begin{align}
\Lambda_{1}(\vec \rho) &\sim 1+2\delta \rho^- -2\delta^2\rho^-\rho^+ ,\\ 
\Lambda_{2}(\vec \rho) &\sim -1 - 2\delta \rho^+ + 2\delta^2\rho^-\rho^+.
\end{align}
\end{subequations}

Although the solutions obtained by the matched asymptotic expansion are expected to be accurate up to first order in $\delta$, we have included $O(\delta^2)$ terms in these expressions for comparison with the eigenvalues obtained with the Rost method below. As expected, the $O(\delta)$ terms in these equations indicate that the speeds of particles moving to the right and to the left are increased by the presence of particles moving in the opposite direction.

For the model \eqref{decomp_num0} obtained using Rost's method, we have that
\begin{equation*}
   {\bf M}=\begin{pmatrix}1+\frac{\partial Z}{\partial \rho^+}&\frac{\partial Z}{\partial \rho^-}\\-\frac{\partial Z}{\partial \rho^+}& -1 - \frac{\partial Z}{\partial \rho^-}
    \end{pmatrix},
\end{equation*}
where 
\begin{equation*}
Z=\frac{2N\epsilon \rho^+ \rho^- }{1-N\epsilon \rho}.
\end{equation*}
In this case, the eigenvalues take a remarkably simple form:
\begin{align} \label{vaps_rost}
\Lambda_{1}(\vec \rho) = \frac{1+\epsilon N\rho^--N\epsilon \rho^+}{1-N\epsilon \rho} \; > \; 
\Lambda_{2}(\vec \rho) = -\left(\frac{1+\epsilon N\rho^+-N\epsilon \rho^-}{1-N\epsilon \rho}\right).
\end{align}

In order to compare with \eqref{evalsMAE}, we use $ \epsilon N \sim \delta$ (for large $N$) in \eqref{vaps_rost} and expand in powers of $\delta$ to find that
\begin{subequations} \label{evalsRost}
\begin{align}
\Lambda_{1}(\vec \rho) &\sim 1+2\delta\rho^- +\delta^2\rho^-\rho ,\\ 
\Lambda_{2}(\vec \rho) &\sim  -1 - 2\delta \rho^+ - \delta^2\rho^+\rho.
\end{align}
\end{subequations}
Thus the wave speeds \eqref{evalsRost} obtained by Rost's method, which was based on the assumption that $N$ is large but with no restrictions on $\delta$, have both $O(\delta)$ and $O(\delta^2)$ terms that contribute to speeds greater than those of the linear noninteracting particles system. The eigenvalues \eqref{evalsMAE} of the system obtained by matched asymptotic expansion were derived on the assumption that $\delta$ is small, and only terms up to $O(\delta)$ are valid. Indeed, the $O(\delta^2)$ contributions to \eqref{evalsMAE} cause a decrease in the predicted wave speeds, in contrast to the effect predicted by \eqref{evalsRost}. This is a further explanation for the qualitative differences in wave shifts observed in Figure \ref{fig:means}.

We note that, since Rost's model \eqref{decomp_num0} is exact for any $\delta$ in the limit of $N\to \infty$ and space-independent turning rates $\lambda^\pm$, we can use its solution and metrics derived from it, such as the speed of the waves \eqref{vaps_rost} as benchmarks to compare to our matched asymptotics approximation. We know that the latter is valid for $\delta$ small, but how far can we push it? For example, comparing \eqref{evalsMAE} with \eqref{evalsRost} we can quantify the  error in the speed of the waves of the asymptotic approximation to be $3 \delta^2 \rho^- \rho^+ + O(\delta ^3)$. 
Although the benchmark Rost's solution is not valid for the general case of varying turning rates or smaller population sizes, we expect these effects to be less important than the occupancy $\delta$.

%%%%%%%%%%%%%%%%%%%%%%%%%%%%%%%%%%%%%%%%%%%%%%
\subsubsection{Transient solutions with tumbling and bias} \label{sec:Dis_Trans}
%%%%%%%%%%%%%%%%%%%%%%%%%%%%%%%%%%%%%%%%%%%%%%

We now present two examples of transient solutions with biased turning rates that depend on the spatial coordinate. Since the kinetic model for rods via Rost's method requires constant turning rates, in this section we only use the nonlinear kinetic model via matched asymptotic expansions \eqref{both-w} to compare with the particle system. Fig.~\ref{Fig:Twaves1} shows the results for the Case 1 parameters in \eqref{signal1} with an excluded fraction of $10\%$, and  Fig.~\ref{Fig:Twaves2} corresponds to the Case 2 \eqref{signal2} with an excluded fraction of $20\%$. In both figures we plot the right-moving, left-moving and total densities $\rho^+$, $\rho^-$ and $\rho$ respectively at three instances in time. As in the previous figures, we compare the PDE solutions for noninteracting (red dashed lines) and hard-core particles (solid blue lines), together with the simulations of the particle system (blue circles).  

\def \scl {1.2}
\begin{figure*}
\unitlength=1cm
\psfrag{t1}[][][\scl]{$t = 0.1$} \psfrag{t2}[][][\scl]{$t = 0.3$} \psfrag{t3}[][][\scl]{$t = 0.5$}
\psfrag{r1}[][][\scl]{$\rho^+$} \psfrag{r2}[][][\scl]{$\rho^-$} \psfrag{r3}[][][\scl]{$\rho$} \psfrag{x}[][][\scl]{$x$}
\psfrag{(a)}[][][\scl]{$(a)$} \psfrag{(b)}[][][\scl]{$(b)$} \psfrag{(c)}[][][\scl]{$(c)$}
\psfrag{(d)}[][][\scl]{$(d)$} \psfrag{(e)}[][][\scl]{$(e)$} \psfrag{(f)}[][][\scl]{$(f)$}
\psfrag{(g)}[][][\scl]{$(g)$} \psfrag{(h)}[][][\scl]{$(h)$} \psfrag{(i)}[][][\scl]{$(i)$}
\includegraphics[width = 0.9\textwidth]{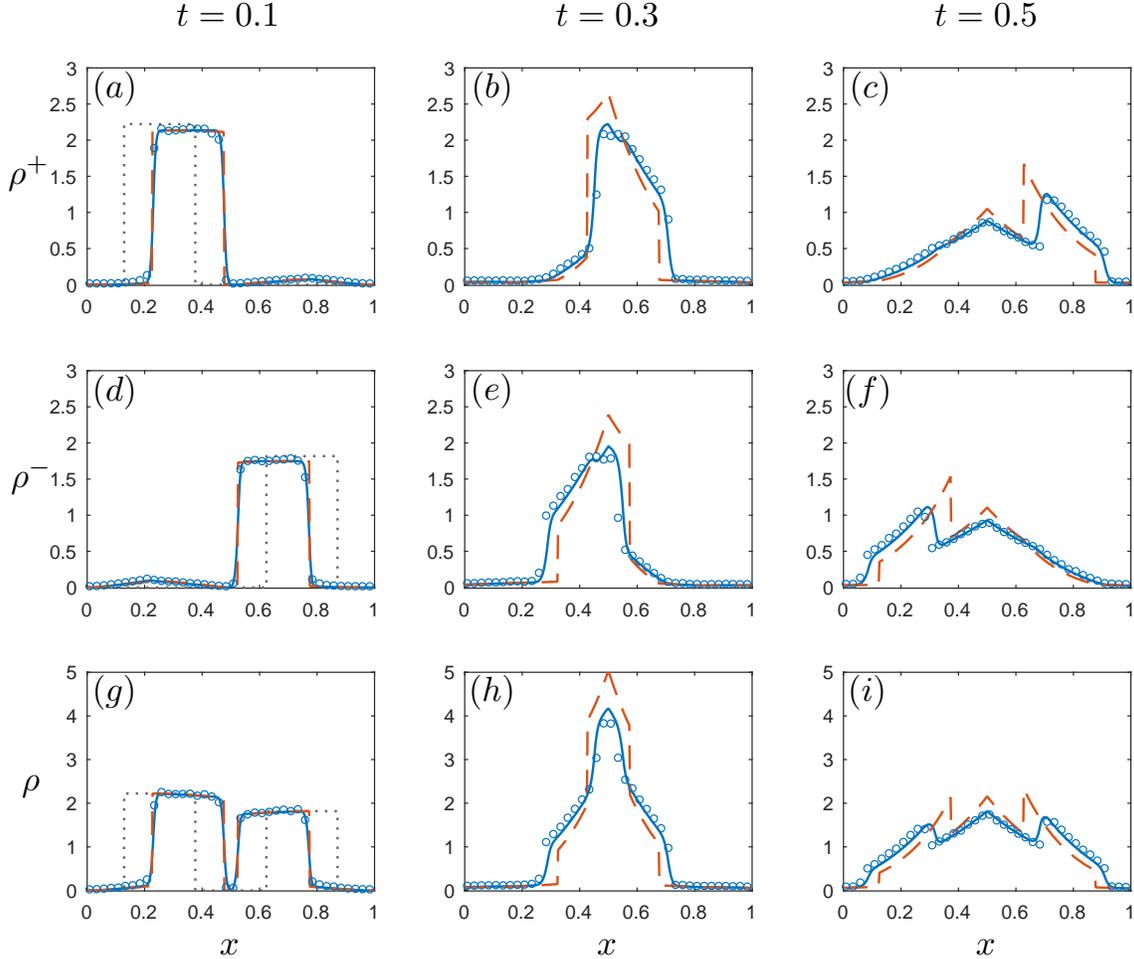}	
\caption{\label{Fig:Twaves1} Transient marginal densities $\rho(x,t)$ and $\rho^{\pm}$ at various times $t > 0$, corresponding to $S_1$. Dotted grey line the initial travelling bands $t = 0$ with uniformly distributed initial data and $N = 100$ total. Dashed red line is solution $\rho(x,t)$ and $\rho^{\pm}$ of \eqref{pde1} for noninteracting particles $(\epsilon = 0)$. Solid blue line is solution $\rho(x,t)$ and $\rho^{\pm}$ of \eqref{both-w} for hard rod particles $(\epsilon = 0.001)$. Blue circles for $\epsilon = 0.001$ computed from $2.5 \times 10^3$ realisations of the KMC method.}
\end{figure*}
\def \scl {1.2}
\begin{figure*}
\unitlength=1cm
\psfrag{t1}[][][\scl]{$t = 0.1$} \psfrag{t2}[][][\scl]{$t = 0.3$} \psfrag{t3}[][][\scl]{$t = 0.8$}
\psfrag{r1}[][][\scl]{$\rho^+$} \psfrag{r2}[][][\scl]{$\rho^-$} \psfrag{r3}[][][\scl]{$\rho$} \psfrag{x}[][][\scl]{$x$}
\psfrag{(a)}[][][\scl]{$(a)$} \psfrag{(b)}[][][\scl]{$(b)$} \psfrag{(c)}[][][\scl]{$(c)$}
\psfrag{(d)}[][][\scl]{$(d)$} \psfrag{(e)}[][][\scl]{$(e)$} \psfrag{(f)}[][][\scl]{$(f)$}
\psfrag{(g)}[][][\scl]{$(g)$} \psfrag{(h)}[][][\scl]{$(h)$} \psfrag{(i)}[][][\scl]{$(i)$}
\includegraphics[width = 0.9\textwidth]{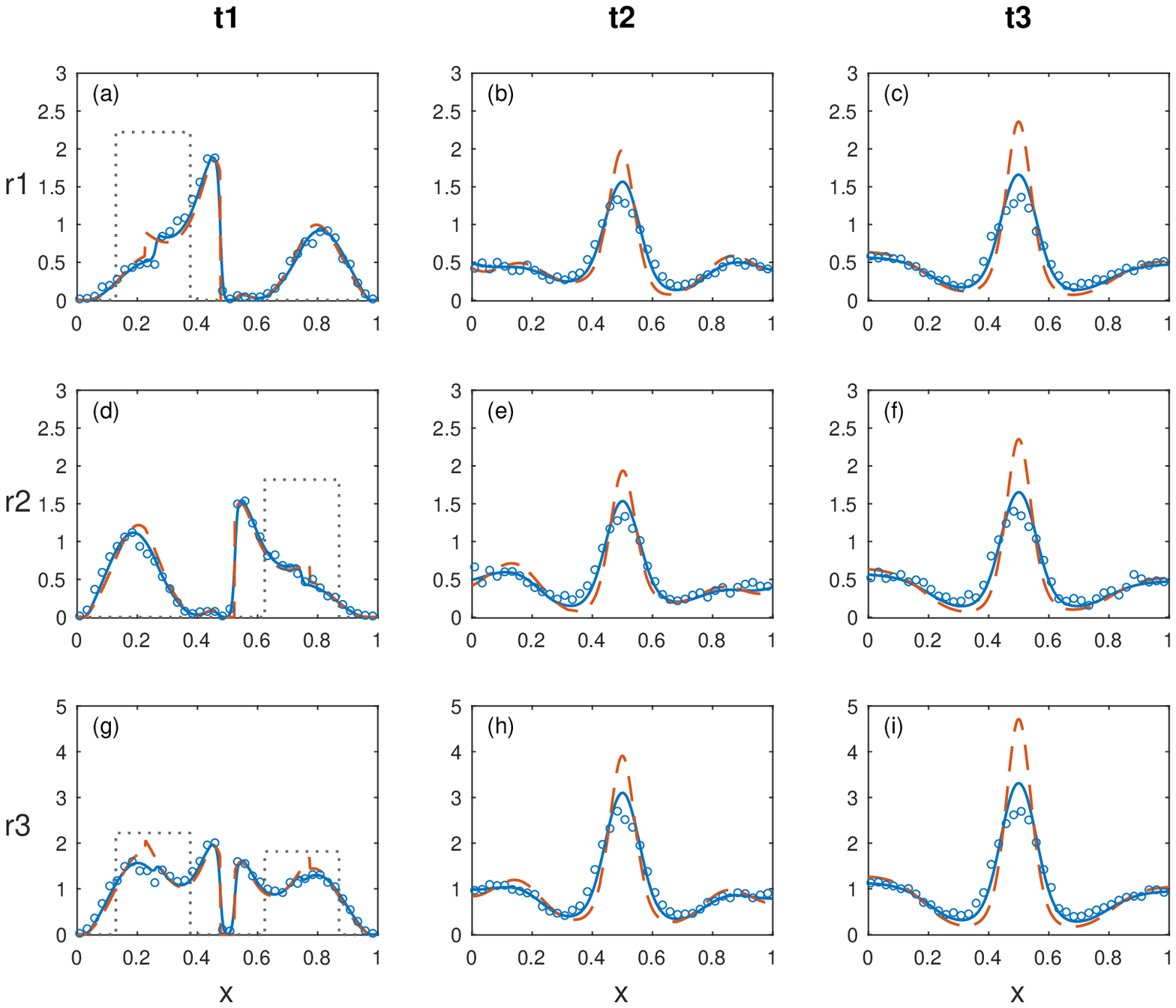}	
\caption{\label{Fig:Twaves2} Transient marginal densities $\rho(x,t)$ and $\rho^{\pm}$ at various times $t > 0$, corresponding to $S_2$. Dotted grey line the initial travelling bands $t = 0$ with uniformly distributed initial data and $N = 100$ total. Dashed red line is solution $\rho(x,t)$ and $\rho^{\pm}$ of \eqref{pde1} for noninteracting particles $(\epsilon = 0)$. Solid blue line is solution $\rho(x,t)$ and $\rho^{\pm}$ of \eqref{both-w} for hard rod particles $(\epsilon = 0.002)$. Blue circles for $\epsilon = 0.002$ computed from $10^2$ realisations of the KMC method.}
\end{figure*}

In both examples we see a good match between the nonlinear kinetic model (solid blue lines) and the particle simulations (blue circles) even with a filled fraction of $20\%$ (see Fig.~\ref{Fig:Twaves2}). A slight discrepancy between the two can be seen in Fig.~\ref{Fig:Twaves2} for times $t\ge 0.3$ in the centre of the domain. Here, the signal gradient is high and that results in a local occupied fraction much greater than $20\%$ and therefore the kinetic model does not perform as well. In principle we could go back to the method of matched asymptotic expansions and consider higher order terms. This could lead to another correction term being added to \eqref{both-w}. The solutions of this augmented kinetic model might come more into line with the simulations apparent at time $t = 0.8$, Fig.~\ref{Fig:Twaves2}(i).
Regarding the effect of the interactions, we notice that in the noninteracting particles case the solutions preserve the discontinuities of the initial condition, whereas the solutions of the nonlinear model appear to smoothed out in time (see in particular the middle and right columns in Fig.~\ref{Fig:Twaves1}). In other words, the excluded-volume interactions seem to act as a diffusion term, which also competes with the external bias (as seen with a reduced peak around the maximum of the signal $S$).

We comment on the closeness of the PDE solutions for noninteracting (red dashed lines) and finite-size particles (blue solid lines) for the initial time $t = 0.1$ in Figs.~\ref{Fig:Twaves1} and \ref{Fig:Twaves2}. Considering that $10\%$ and $20\%$ respectively of the domain is occupied by particles of a finite size one could expect the difference to be greater. We can explain the closeness in this manner: in the beginning there is a train of particles coming in from the left and another train coming in from the right. Whether the particle has a finite size or not they are still subject to the same Poisson processes with rates $\lambda(x_i,v_i)$. Before the two trains collide roughly at the centre of the domain this is the dominating process that dictates the behaviour of the particles. In addition to this there would be a number of instances in time where a finite-size particle after having previously jumped back independently may again switch velocity only after colliding with a neighbour (since the random walk is biased towards the centre jump back collisions are less likely to occur). Up until the two trains meet this can be viewed like a secondary effect, the overall group velocities would be close to $c$ and $-c$ respectively. Only after the two trains meet at the centre does the switching of velocities due to collisions between finite-size particles become more significant. The overall group velocities become disrupted and the difference between the linear and nonlinear PDE solutions become more noticeable (see for example the middle column in Fig.~\ref{Fig:Twaves1}).

In the second example, by time $t = 0.3$ the kinetic model has already moved into the diffusion mode (middle column in  Fig.~\ref{Fig:Twaves2}). This has come about because the relatively high baseline turning frequency $\lambda_0 = 16$ is to a certain extent mimicking Brownian motion. This is contrary to the first case, where the relatively low $\lambda_0 = 2.5$ allows the kinetic waves to continue for times $t \geq 0.3$ (see Fig.~\ref{Fig:Twaves1}). Yet, by $t = 0.5$ in the first example we observe that the kinetic model begins to converge to the stationary solution (compare Fig.~\ref{Fig:Twaves1}(i) with Fig.~\ref{Fig:MH}(a)). In the second example, in contrast to the linear model, the nonlinear kinetic model is already very close to the stationary solution at time $t = 0.3$, since we see little change in the solution between times $t = 0.3$ and $t = 0.8$ (see Figs.~\ref{Fig:Twaves2}(h) and (i)). The reason for this is that the collisions between hard-core particles accelerate the convergence to the diffusive regime and the stationary solution, which is consistent with the enhanced diffusion coefficient of the diffusion limit \eqref{diffusion_limit}.

%%%%%%%%%%%%%%%%%%%%%%%%%%%%%%%%%%%%%%%%%%%%%%
\section{Summary and discussion} \label{sec:conclusions}
%%%%%%%%%%%%%%%%%%%%%%%%%%%%%%%%%%%%%%%%%%%%%%

In this paper we have considered a velocity jump process with excluded-volume interactions. In particular, starting with a system of $N$ hard-core rod particles that switch their velocities with elastic collisions in one dimension while the constant speed is always preserved, we have derived a nonlinear kinetic model using two different approaches. The first approach, based on matched asymptotic expansions, is systematic and hence does not rely on a closure assumption. It is valid in the limit of small but finite particle occupied fraction and in the presence of external signals (leading to a bias in the tumbling rates). The second method, based on a compression method by Rost \cite{rost1984diffusion}, does not have the limitation of a small occupied fraction but assumes that the tumbling rates are constant (independent of the spatial variable). Therefore the latter can only capture linear chemical gradients. By considering a parabolic scaling, we have obtained the diffusive limit of the kinetic model and seen it agrees with the single-file diffusion model that one obtains starting from a set of Brownian hard rods \cite{bruna2014diffusion}. 

Excluded-volume interactions emerge in the kinetic model  as a nonlinear transport term, proportional to the density of particles moving in the opposite direction. We have validated our nonlinear model with numerical simulations, comparing its solutions with the corresponding noninteracting linear model as well as stochastic simulations of the underlying particle system. We have considered transient and stationary solutions under different tumbling, external bias, and excluded-volume scenarios.

The method of matched asymptotic expansions had previously been used in the context of Brownian particles or parabolic PDEs. Here we have shown it also generalises to hyperbolic systems (we note that in \cite{franz2016hard} the method had been quoted in the context of a velocity-jump process for hard disks, but it was not actually used to solve the two-particle density problem). It would be interesting to see if the method can be used in higher dimensions to derive the kinetic model in a systematic way and compare it with the results in \cite{franz2016hard}.

The method was implemented here in its simplest setting, hard-core identical rods influenced by a fixed signal in the domain similar to an external potential. If the signal is thought of as a chemotactic signal, a natural extension would be to consider a more realistic chemotaxis model to study the interplay between the signal concentration and the finite-size effects. In the context of Brownian hard-core particles, this was considered in \cite{Wilson:2018fg}. 

Another interesting direction would be to generalise the interactions between particles and either consider a mixed hard and soft interacting potential between particles (similar to the ones considered in \cite{bodnar2005derivation}), which may be more realistic in the context of biological applications than a bare hard-core interaction or a softer interaction that allows particles to overlap each other. 

Finally, an important aspect when modelling cell biology is to allow for multiple subpopulations and/or changing numbers of particles to account for processes such as cell growth, proliferation, and phenotypic switching. Models for cell growth incorporate the cell size as a variable, and the total number of cells is also a variable due to death and birth events (see for example \cite{zaidi2015} and references therein). One could incorporate such ideas to include cell movement in the context of our work, keeping in mind that cell growth implies that the total excluded volume would also become a variable. This makes the derivation more complicated but also potentially more interesting, as keeping track of excluded volume interactions explicitly would mean that at some point growth and proliferation are not possible anymore (in a similar way that models for cell chemotaxis that ignore excluded volume can result in finite-time blow up, something that can be prevented when incorporating excluded-volume interactions into the model \cite{Burger:2006fq}). It would be also interesting to see if the problems associated with lattice based models of cell movement with proliferation \cite{simpson2014236} can be avoided with an off-lattice approach.

\begin{acknowledgments}
The authors would like to thank Jos{\'e} A. Carrillo from Imperial College London for suggesting to consider the diffusive limit presented in Section~\ref{sec:difflimit} and Nich Hale from Stellenbosch University for his support using Chebfun to solve the kinetic model. MB thanks the Royal Society for a University Research Fellowship   (grant URF$\backslash$R1$\backslash$180040).
\end{acknowledgments}

\appendix
\section{Integrated equation} \label{sec:integration}
\begin{figure}[tbh]
\includegraphics[width = .4\textwidth]{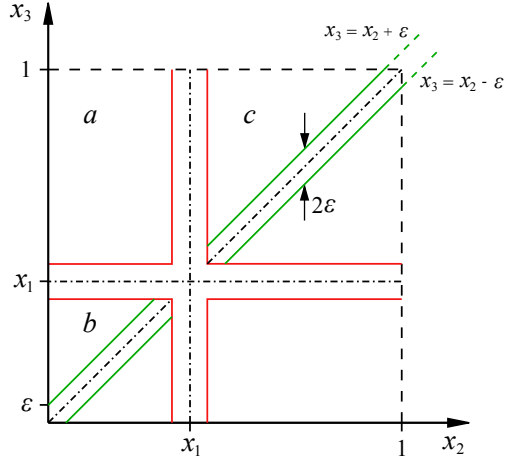}
\caption{\label{fig:integration} Sketch of $\Omega_\epsilon^3(x_1)$ and the excluded regions.} 
\end{figure}

Here we detail the steps to arrive at the integrated equation \eqref{integ1} from the original transport equation \eqref{Ndim_equation}. We first assume that $N=3$, and so \eqref{Ndim_equation} reads
\begin{equation}
	\label{3dim_equation}
	\frac{\partial P}{\partial t} + v_1 \frac{\partial P}{\partial x_1} + v_2 \frac{\partial P}{\partial x_2} + v_3 \frac{\partial P}{\partial x_3}  + \sum_{i=1}^3 \left[\lambda(x_i,v_i)P(\vec x,\vec v,t)-\lambda(x_i,-v_i)P(\vec x,s_i \vec v,t) \right] = 0.
\end{equation}
To obtain an equation for the marginal density function $p(x_1,v_1,t)$ \eqref{marginal1}, we need to integrate \eqref{3dim_equation} in the spatial domain available for $x_2$ and $x_3$ given that  the first particle is at position $x_1$. This is given by (see Fig.~\ref{fig:integration})
\begin{equation}
	\label{domain_N3}
	\Omega^3_\epsilon(x_1) = \left \{ (x_2, x_3) \in \Omega^2 : | x_1-x_2| \ge \epsilon, | x_1-x_3| \ge \epsilon, | x_2-x_3| \ge \epsilon  \right \}.
\end{equation}
$\Omega^3_\epsilon(x_1)$ has six disconnected regions provided that $2 \epsilon < x_1 < 1-2 \epsilon$, four if $x_1$ is between $\epsilon$ and $2\epsilon$ away from a wall, and two if it is closer than $\epsilon$ from a wall. 

The integral of the first term in \eqref{3dim_equation} is $\partial _t p(x_1,t)$ from \eqref{marginal1} noting that $\Omega^3_\epsilon(x_1)$ is independent of time. The only terms remaining after integrating the fifth term in \eqref{3dim_equation} are those corresponding to $i=1$; the rest cancel out after summing $v_i$ for $i\ge 2$ over $\{-c, c\}$.

The third and fourth terms in \eqref{3dim_equation} will give the same by the symmetry of $\Omega^3_\epsilon(x_1)$ with respect to $x_2$ and $x_3$. Therefore they give two copies of
\begin{align*}
	\int_{\Omega^3_\epsilon(x_1) \times V^2} &v_2 \frac{\partial P}{\partial x_2} \ud x_2 \ud v_2 \ud x_3  \ud v_3 = \int_{\Omega^3_\epsilon(x_1)\times V } \sum_{v_2 = \pm c} v_2\frac{\partial P}{\partial x_2} \ud x_2 \ud x_3 \ud v_3 \\
	&=  \int_{\Omega_\epsilon^3(x_1,x_2) \times V} \sum_{v_2 = \pm c} v_2 \left[ P\big|^{x_2=1}_{x_2 = x_1+\epsilon} + P\big|^{x_2= x_1-\epsilon}_{x_2 = 0} \right] \ud x_3 \ud v_3 \\
	&=  \sum_{v_2 = \pm c} v_2  \int_{\Omega_\epsilon^3(x_1,x_2) \times V} \left[ - P\big|_{x_2 = x_1+\epsilon} + P\big|_{x_2= x_1-\epsilon}\right] \ud x_3 \ud v_3 = - \sum_{v_2 = \pm c} v_2 P_2 \big|_{x_2=x_1-\epsilon}^{x_2 =x_1+\epsilon}.
\end{align*}
In the last two lines we assume that $\epsilon < x_1 < 1-\epsilon$ so that the red cross in Fig.~\ref{fig:integration} is inside $\Omega^2$; otherwise the terms that would evaluate $x_2 \notin \Omega$ are dropped. The terms in the second line above evaluated at $x_2 = 0,1$ vanish using the boundary condition \eqref{collision_wall}. Finally, the last equality comes from using the definition of $P_2$ \eqref{marginal2}.

We are left with the second term in \eqref{3dim_equation}, which requires care since the domain of integration does depend on $x_1$. The integrals of the regions above and below the diagonal $x_2 = x_3$ are equal by symmetry, so we focus on the upper part. The Leibniz rule reads
\begin{equation*}
	\partial_{x_1} \int_{\Omega_\epsilon^3(x_1)} P \ud x_2 \ud x_3 = \int_{\Omega_\epsilon^3(x_1)} \partial_{x_1} P \ud x_2 \ud x_3  + \int_{\partial \Omega_\epsilon^3(x_1)} P  ({\bf u}\cdot {\bf \hat n} ) \ud s,
\end{equation*}
where $\bf u$ is the ``velocity'' of the boundary with respect to $x_1$ and $\bf \hat n$ is the unit outward normal vector to the boundary $\partial \Omega_\epsilon^3(x_1)$ (these are 2-dimensional vectors in the $x_2 x_3$ plane). The only boundaries of $\Omega_\epsilon^3(x_1)$ that move with $x_1$ are the vertical and horizontal lines $x_2 = x_1 \pm \epsilon$ and $x_3 = x_1 \pm \epsilon$ (depicted in red in Fig.~\ref{fig:integration}), for which ${\bf u} = (1,0)$ and ${\bf u} = (0, 1)$ respectively. 
Using this for example to integrate in the area marked as $a$ in Fig.~\ref{fig:integration}
\begin{align*}
	\int_{\partial a}  P  ({\bf u}\cdot {\bf \hat n} ) \ud s &= \int_0^{x_1-\epsilon} P(x_1,x_2, x_1+\epsilon,\vec v,t)(-1) \ud x_2 + \int_{x_1+\epsilon}^1 P(x_1,x_1-\epsilon,x_3,\vec v,t)(+1) \ud x_3  \\
	&= \int_0^{x_1-\epsilon} P(x_1,x_1+\epsilon,x_3,\vec v,t)(-1) \ud x_3 + \int_{x_1+\epsilon}^1 P(x_1,x_1-\epsilon,x_3,\vec v,t)(+1) \ud x_3,
\end{align*}
by relabelling invariance. The other two regions $b$ and $c$ give
\begin{align*}
	\int_{\partial b}  P  ({\bf u}\cdot {\bf \hat n} ) \ud s &=\! \int_0^{x_1-2\epsilon} \! P(x_1,x_2, x_1-\epsilon,\vec v,t)(+1) \ud x_2 \!= \! \int_0^{x_1-2\epsilon} \! P(x_1,x_1-\epsilon,x_3,\vec v,t)(+1) \ud x_3,\\ 
	\int_{\partial c}  P  ({\bf u}\cdot {\bf \hat n} ) \ud s &=\! \int_{x_1+2\epsilon}^1 \! P(x_1, x_1+\epsilon,x_3,\vec v,t)(-1) \ud x_3.
\end{align*}
Again, some of these regions might be empty depending on how close $x_1$ is to the boundary. As $x_1$ moves away from the boundary the excluded area $|\Omega_\epsilon^3(x_1)|$ increases as ``new boundary'' is created, but the expressions above remain valid as long as $P$ is non-singular in the vicinity of the new boundaries.   
Combining these we have 
\begin{equation*} 
\begin{split}
	\frac{1}{2} \int_{\partial \Omega_\epsilon^3(x_1)} P  ({\bf u}\cdot {\bf \hat n} ) \ud s  =  &\int_{\Omega_\epsilon^3(x_1,x_1-\epsilon)}  P(x_1, x_1-\epsilon, x_3, \vec v, t) \ud x_3 \\
	&-\int_{\Omega_\epsilon^3(x_1,x_1+\epsilon)}  P(x_1, x_1+\epsilon, x_3, \vec v, t) \ud x_3,
\end{split}
\end{equation*}
where we recall that $\Omega_\epsilon^3(x_1,x_2)$ is the interval in $\Omega$ available to the third particle if the first and second particles are at $x_1$ and $x_2$, respectively. 
Therefore,
\begin{align*}
	\int_{\Omega^3_\epsilon(x_1) \times V^2} v_1 \frac{\partial P}{\partial x_1} \ud x_2 \ud x_3 \ud v_2  \ud v_3 &= v_1 \partial_{x_1} \int_{\Omega^3_\epsilon(x_1)\times V^2 } P \ud x_2 \ud x_3  \ud v_2  \ud v_3 \\
	&\phantom{= v_1 \partial_{x_1} p} - v_1\int_{V^2} \int_{\partial \Omega_\epsilon^3(x_1)} P  ({\bf u}\cdot {\bf \hat n} ) \ud s  \ud v_2  \ud v_3 \\
	&= v_1 \partial_{x_1} p - 2v_1\sum_{v_2 = \pm c} \int_{\Omega_\epsilon^3(x_1,x_1-\epsilon)\times V} \! P(x_1, x_1-\epsilon, x_3, \vec v, t) \ud x_3 \ud v_3 \\
	&\phantom{= v_1 \partial_{x_1} p} + 2v_1\sum_{v_2 = \pm c} \int_{\Omega_\epsilon^3(x_1,x_1+\epsilon) \times V} \! P(x_1, x_1+\epsilon, x_3, \vec v, t) \ud x_3 \ud v_3 \\
	&= v_1 \partial_{x_1} p + 2 v_1 \sum_{v_2 = \pm c}  P_2 \big|_{x_2=x_1-\epsilon}^{x_2 =x_1+\epsilon}.
\end{align*}
Adding all the terms we obtain \eqref{integ1} with $N=3$. The general $N$ case is obtained by noting that adding further dimensions reduces to copies of the terms involving $P_2$ (by using the invariance of label permutations to relabel).

%\bibliography{paper_biblio}% Produces the bibliography via BibTeX.

\providecommand{\noopsort}[1]{}\providecommand{\singleletter}[1]{#1}%

\end{document}